\documentclass[journal]{IEEEtran}
\usepackage{cite}
\usepackage{times}
\usepackage{epsfig}
\usepackage{graphicx}
\usepackage{amsmath}
\usepackage{amssymb}
\usepackage{epstopdf}
\usepackage{subfigure}
\usepackage{url}
\usepackage{multirow}
\usepackage{booktabs}
\usepackage{tabularx}
\usepackage[noend]{algpseudocode}
\usepackage{algorithmicx,algorithm}

\begin{document}

\title{Fast and High-Performance Learned Image Compression With  Improved Checkerboard Context Model, Deformable Residual Module, and Knowledge Distillation}
\author{Haisheng~Fu,
        Feng~Liang,
        Jie~Liang,
        Yongqiang~Wang,
        Guohe~Zhang,
        Jingning~Han
\thanks{Haisheng~Fu, Feng~Liang, Yongqiang~Wang and Guohe~Zhang are with the School of Microelectronics, Xi'an Jiaotong University, Xi'an, China.  (e-mails: fhs4118005070@stu.xjtu.edu.cn; fengliang@xjtu.edu.cn; wangyq0901@stu.xjtu.edu.cn; zhangguohe@xjtu.edu.cn) (Corresponding authors: Feng Liang).}% <-this % stops a space
\thanks{Jie Liang is with the School of Engineering Science, Simon Fraser University, Canada (e-mails: jiel@sfu.ca).}% <-this % stops a space
\thanks{Jingning~Han is with the Google Inc. (e-mail: jingning@google.com).}
\thanks{This work was supported by the National Natural Science Foundation of China (No. 61474093), the Natural Science Foundation of Shaanxi Province, China (No. 2020JM-006), the Natural Sciences and Engineering Research Council of Canada (RGPIN-2020-04525), China Scholarship Council, and Google Chrome University Research Program.} }

%\markboth{Submitted to IEEE Transactions on Neural Networks and Learning Systems}%
\markboth{Submitted to Trans. journal}%
%\markboth{IEEE Transactions on Multimedia}%
{Fu \MakeLowercase{\textit{et al.}}: Fast and High-Performance Learned Image Compression With  Improved Checkerboard Context Model, Deformable Residual Module, and Knowledge Distillation}

\maketitle


\begin{abstract}
Deep learning-based image compression has made great progresses recently. However, many leading schemes use serial context-adaptive entropy model to improve the rate-distortion (R-D) performance, which is very slow. In addition, the complexities of the encoding and decoding networks are quite high and not suitable for many practical applications. In this paper, we introduce four techniques to balance the trade-off between the  complexity  and performance. We are the first to introduce deformable convolutional module in compression framework, which can remove more redundancies in the input image, thereby enhancing compression performance.  Second, we design an improved checkerboard context model with two separate distribution parameter estimation networks and different probability models, which enables parallel decoding without sacrificing the performance compared to the sequential context-adaptive model. Third, we develop a three-step knowledge distillation and training scheme to achieve different trade-offs between the complexity and the performance of the decoder network, which transfers both the final and intermediate results of the teacher network to the student network to help its training. Fourth, we introduce $L_{1}$ regularization to make the numerical values of the latent representation more sparse. Then we only encode non-zero channels in the encoding and decoding process, which can greatly reduce the encoding and decoding time. Experiments show that compared to the state-of-the-art learned image coding scheme, our method can be about 20 times faster in encoding and 70-90 times faster in decoding, and our R-D performance is also $2.3 \%$ higher. Our method  outperforms the traditional approach in H.266/VVC-intra (4:4:4) and some  leading learned schemes in terms of PSNR and MS-SSIM metrics when testing on Kodak and Tecnick-40 datasets.

\end{abstract}

%%%%%%%%% BODY TEXT

\section{Introduction}
\label{sec:intro}

Recently deep learning has been successfully applied to the field of image compression with very impressive results. The main components of classical image compression standards, e.g., JPEG \cite{JPEG}, JPEG 2000 \cite{JPEG2000}, BPG (intra-coding of H.265/HEVC) \cite{BPG}, and H.266/VVC \cite{VVC}, include linear transform, quantization, and entropy coding. In the end-to-end learning-based framework, these components have been re-designed carefully.

\begin{figure}[!t]
	\centering
		\includegraphics[scale=0.52]{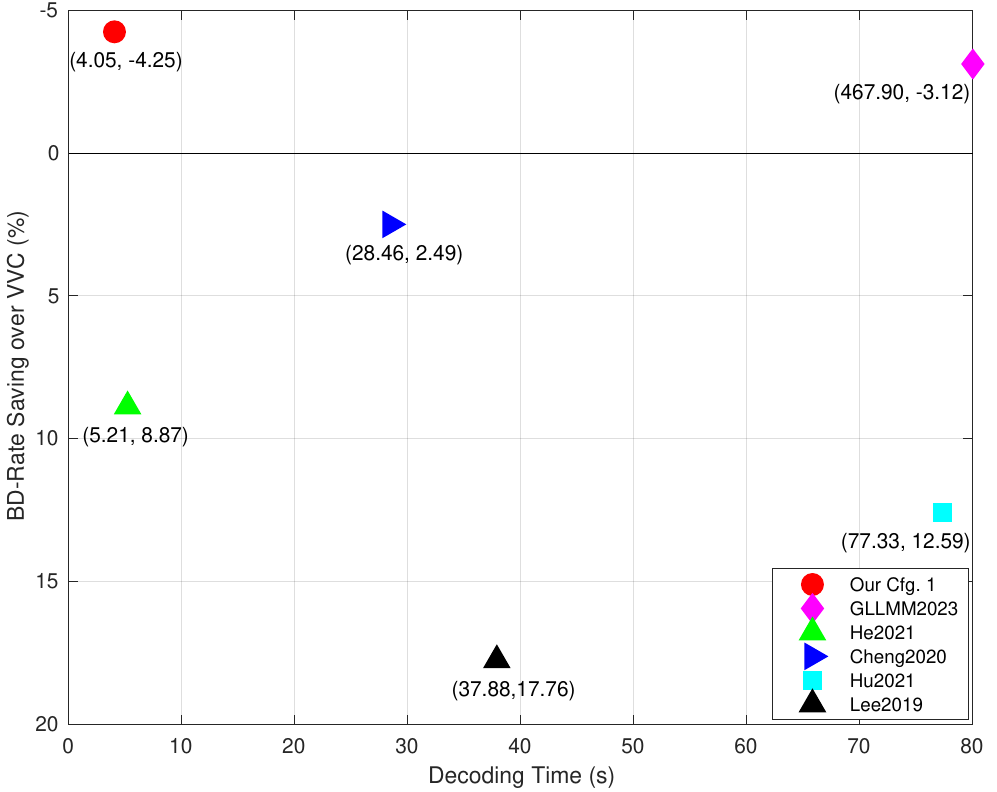}
	\caption{The decoding time and BD-Rate saving over H.266/VVC of different methods for the Kodak dataset. The upper-left corner has better result. The large decoding time of GLLMM \cite{GLLMM} is written explicitly in the bracket.}
	\label{comp_time_bdrate}
\end{figure}

\begin{figure*}[!htp]
	\centering
		\includegraphics[scale=0.6]{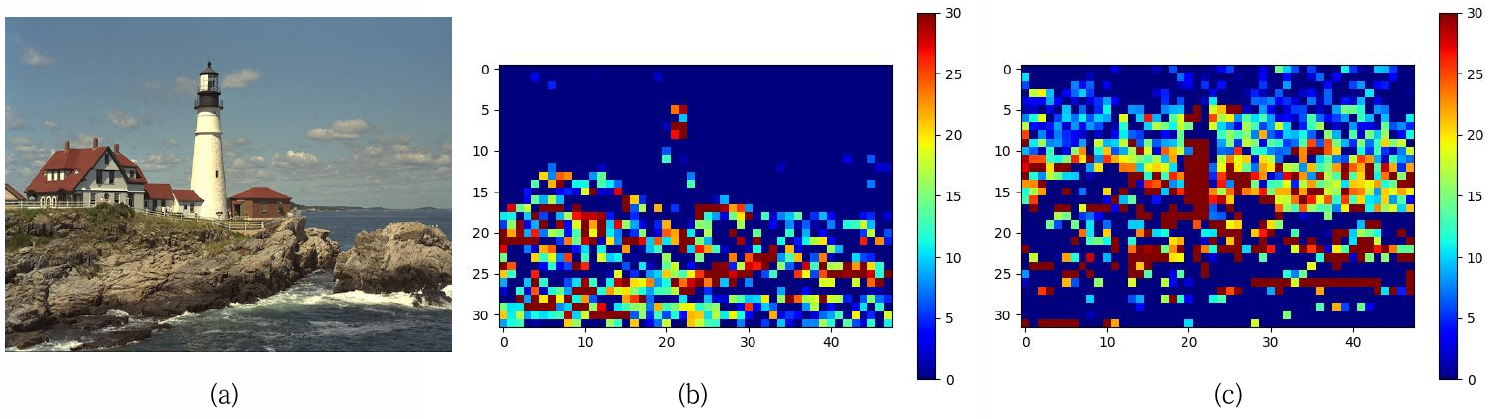}
	\caption{(a) An original image in the Kodak dataset. (b) Illustration of the average value of latent representations with $L_{1}$ regularization. (c) Illustration of the average value of latent representations without $L_{1}$ regularization.}
	\label{fig:L1_norm}
\end{figure*}

In the transform part, various deep learning-based networks have been developed to extract compact latent representations of the input image, such as residual blocks \cite{Chen_TNNLS_2022, FU_2020, Asymmetric_Fu}, attention modules \cite{chen2021, Li_Nonlocal_entropy}, invertible structures \cite{xie2021enhanced}, or transformer blocks \cite{zhu2022transformerbased, Zou_2022_CVPR}. Although these structures significantly improve the rate-distortion (RD) performance, their complexity of the networks is usually quite high.

In the quantization part, since learning-based approach requires all components of the codec to be differentiable, but the traditional quantization is not differentiable, different technologies have been proposed to alleviate this problem. For example, in \cite{Variational, Joint, cheng2020, Asymmetric_Fu, GLLMM}, the quantization is implemented by adding uniform noise to the latent representation during training, and the rounding operation is used during inference.

For the entropy coding part, the application of the serial context-adaptive entropy model significantly improves the rate-distortion (R-D) performance, in which hyperpriors and autoregressive models are jointly utilized to capture the spatial redundancy of the latent representations. However, these methods cannot be accelerated in the decoding process by parallel computing devices, such as FPGA or GPU, making them not suitable for practical applications.

Some recent works using serial context-adaptive entropy model can even outperform the best traditional image standards (i.e. VVC intra coding) in terms of PSNR \cite{Lee_2021, GLLMM, xie2021enhanced}. In particular, the scheme in \cite{GLLMM} represents the current state of the art, where the latent representations are assumed to follow the Gaussian-Laplacian-Logistic mixture model (GLLMM). However, its complexity is quite high.

In this paper, we first propose the deformable residual module to improve the image compression performance. Next, we propose three techniques to reduce the model size and decoding complexity of learned image compression methods while maintaining competitive R-D performance. The main contributions of this paper are summarized as follows:

\begin{itemize}

\item We are the first to propose the deformable residual module (DRM), which combines the deformable convolution (\cite{Defor_Conv_V1})  and residual block \cite{resblock}. The proposed deformable residual module (DRM) can expand the receptive field and is easier to obtain global information. The DRM can further capture and reduce spatial correlation of the latent representations and improve the compression performance.

\item Second, we propose an improved checkerboard context model, which divides the latents into two subsets via a checkerboard pattern, and each of them can be processed in parallel, thereby significantly speeding up the decoding. It uses two different networks to estimate the distribution parameters of the two  subsets. It also only employs the more powerful GLLMM model in the first subset, because it does not use context model. The second subset still use the simpler Gaussian mixture model (GMM), without affecting its R-D performance.

\item Third, we develop a three-step knowledge distillation scheme to achieve different trade-offs between the performance and complexity for the decoder network. The concept of knowledge distillation was first proposed in \cite{hinton2015distilling}, where a lightweight student network is trained to learn the Softmax outputs of a trained and complex teacher model. In our scheme, the student decoder network is first chosen to have the same architecture as the teacher network. We jointly train them to transfer important prior information from the teacher decoder network to the student decoder network to improve its performance. We next use different distillation technologies to reduce the complexity of the student network, such as removing some modules and reducing the number of filters.

\item Fourth, we introduce $L_{1}$ regularization to make the numerical values of the latent representation sparser (as shown in Fig. \ref{fig:L1_norm}), increasing the number of zero elements in the latent representation. Then, in the encoding and decoding process, we only encode non-zero channels, which significantly reduces the encoding and decoding time without sacrificing coding performance.

\end{itemize}

Experiment results using the Kodak and Tecnick-40 datasets show that compared to the state-of-the-art learned image coding scheme in \cite{GLLMM}, our method can be about 20 times faster in encoding and 70-90 times faster in decoding, and our R-D performance is  $2.3 \%$ higher. Our method also outperforms the latest traditional approach in H.266/VVC-intra (4:4:4) and other leading learned schemes such as \cite{cheng2020} in both PSNR and MS-SSIM metrics. The decoding time and BD-Rate comparison with VVC of some methods are reported in Fig. \ref{comp_time_bdrate}.

\begin{figure*}[!thp]
	\centering
		\includegraphics[scale=0.5]{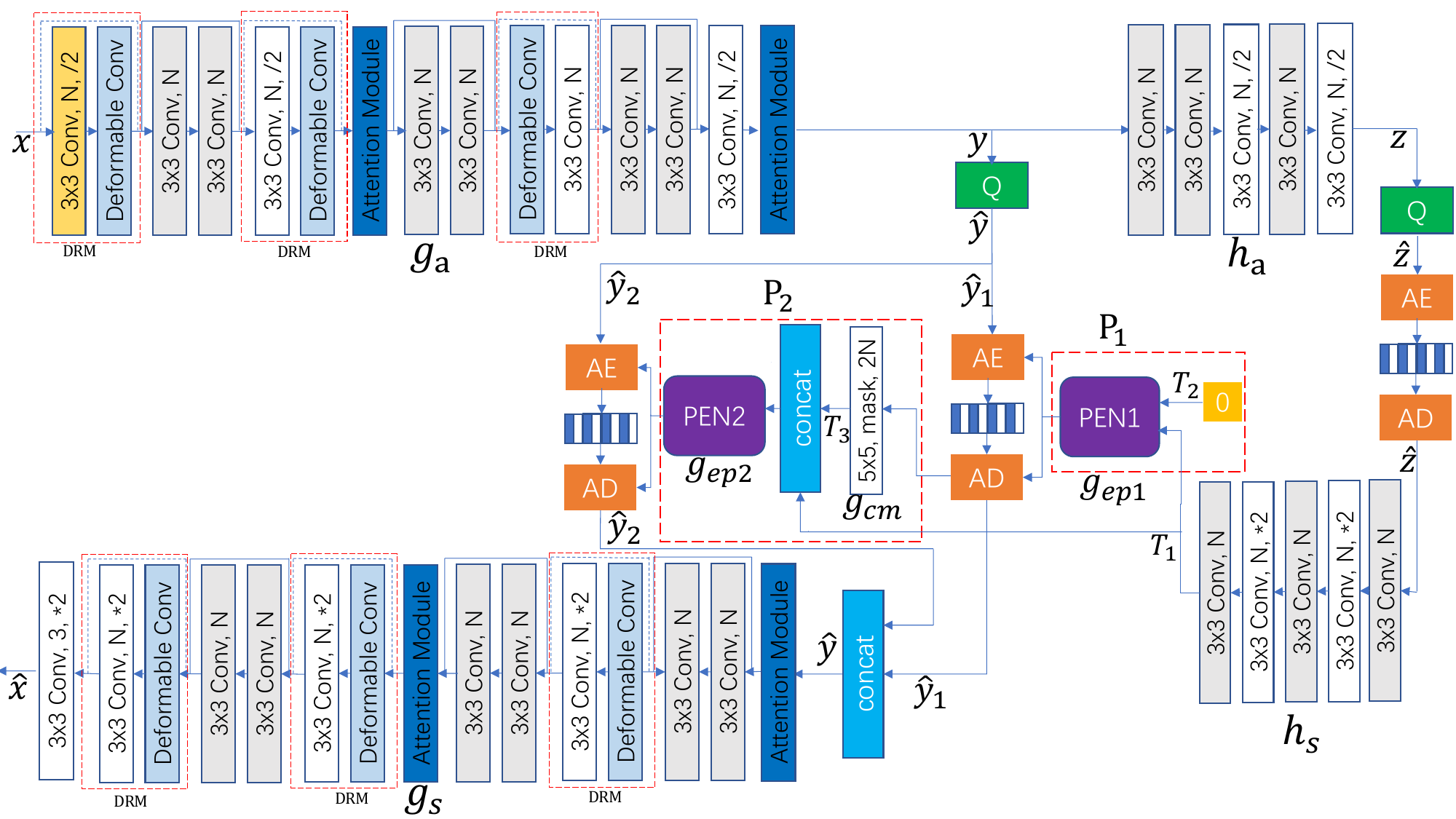}
	\caption{The architecture of the proposed learned image compression scheme. The decoder will be further distilled in Fig. \ref{distillation_framework_model}. $G$ and $IG$ represent generalized divisive normalization (GDN) and inverse GDN  (IGDN). $\uparrow$ and $\downarrow$ denote the up/down-sampling operators. $3 \times 3$ is the convolution size. $AE$ and $AD$ stand for arithmetic encoder and decoder. $L$ represents leaky ReLU activation function. The dotted lines represent the shortcut connection with changed tensor size.}
	\label{whole_networkstructure}
\end{figure*}

\section{Related Work}

\textbf{Context Models.} Most learned image compression methods are based on the autoencoder architecture to extract the compact latent representation of the image \cite{end_to_end}. An autoregressive model is usually used to predict latents from their causal context. In \cite{Variational,Joint}, a hyperprior network is introduced to learn some side information to correct the context-based predictions. The data from the context model and the hyper network are then combined to learn the probability distributions of the quantized latents, and guide the entropy coding. In \cite{Variational,Joint}, simple Gaussian models are used. In \cite{cheng2020,GLLMM}, Gaussian Mixture Model (GMM) and Gaussian-Laplacian-Logistic Mixture Models (GLLMM) are proposed, leading to state-of-the-art performance.

However, serial context models are not friendly to parallel processing during decoding. To address this issue, in \cite{channel}, a channel-wise autoregressive entropy model is proposed to minimize the element-level serial processing in context model. In \cite{He_2022_CVPR}, a spatial-channel contextual adaptive model is proposed to boost the rate-distortion performance without sacrificing running speed. In \cite{He_2021_CVPR}, a checkerboard context model (CCM) is proposed, which divides all data into two groups in a checkerboard pattern to facilitate parallel processing. However, the R-D performance is dropped by 0.2-0.3 dB on the Kodak dataset.

\textbf{Deformable Convolution.} 
Dai et al. \cite{Defor_Conv_V1} first utilize deformable convolution together with the learned offset maps to boost the modeling capability of the neural networks.  The method has achieved better performance than classical convolutions networks in sophisticated vision tasks such as object detection and semantic segmentation. Later, The deformable convolution also has applied in other computer vision tasks, such as action recognition \cite{Klopp_2022_Action}, and video super-resolution\cite{TDAN_CVPR_2020, EDVR_CVPR_2019}. We also note that the deformable convolution has been applied in video compression \cite{FVC}.  The deformable convolution with dynamic kernels is used to better capture more complex non-rigid motion patterns between two consecutive frames, which can  boost
the motion compensation performance and also alleviate the burden for the subsequent residual compression module. In contrast to these prior studies, our work is the first attempt to explore the integration of deformable convolution with the learning-based image compression framework. Considering that there are different components in our learning-based
image compression framework, it is a very challenging task to propose an end-to-end optimized image compression framework by seamlessly incorporating deformable convolution and other modules.

\textbf{Knowledge Distillation.} Knowledge distillation is a method to transfer knowledge from a complex teacher network to a simple student network \cite{Chen_Distillation, Yim_2017_CVPR, gu2022openvocabulary, Li_KD_2022, Yu_KD_2023}. The student model distills knowledge by utilizing gradient descent backpropagation of the distillation loss, which measures the disparity between predictions and soft teacher targets. In \cite{Yang_detection_KD}, the Focal and Global Distillation (FGD) method is proposed to guide the student detector and improves the performance of object detection. In \cite{Tung_KD_2019, Heo_KD_2019},  different knowledge distillation methods are designed for image classification and achieve good performance.  In \cite{GAN_distillation_image_compression}, the knowledge distillation is first introduced to learned image compression. However, it only focuses on visual performance at low bit rates using the Generative Adversarial Network (GAN). Its network architecture does not include the hyper network, and the performance is thus not very good. Moreover, only the prior knowledge of the final output of the teacher network is considered in the distillation. The intermediate results of the teacher network are not distilled.

\section{The Proposed Image Compression Framework}

In this section, we first present the entire architecture of the proposed method. Next, we describe the details of major components, including the improved checkerboard context model, the three-step knowledge distillation of the decoder network, and the corresponding training method.

\begin{figure}
\centering
\subfigure
{\begin{minipage}[t]{0.5\linewidth}
\centering
\includegraphics[scale=0.592]{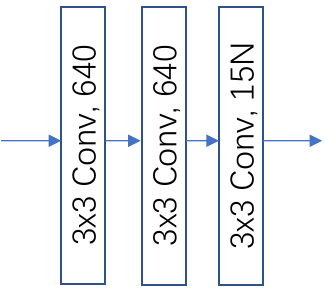}
\label{PEN1_architecture}
\end{minipage}
}%
\subfigure
{\begin{minipage}[t]{0.5\linewidth}
\centering
\includegraphics[scale=0.6]{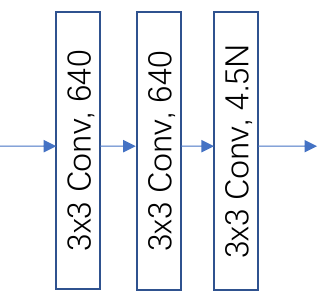}
\label{PEN2_architecture}
\end{minipage}
}%
\centering
  \caption{(a) The detailed architecture of PEN1 network in Fig. \ref{whole_networkstructure}. (b) The detailed architecture of PEN2 network in Fig. \ref{whole_networkstructure}.}
  \label{fig:PEN}
\end{figure}

\subsection{The Overall Architecture of the System} 

\begin{figure}[!thp]
	\centering
		\includegraphics[scale=0.5]{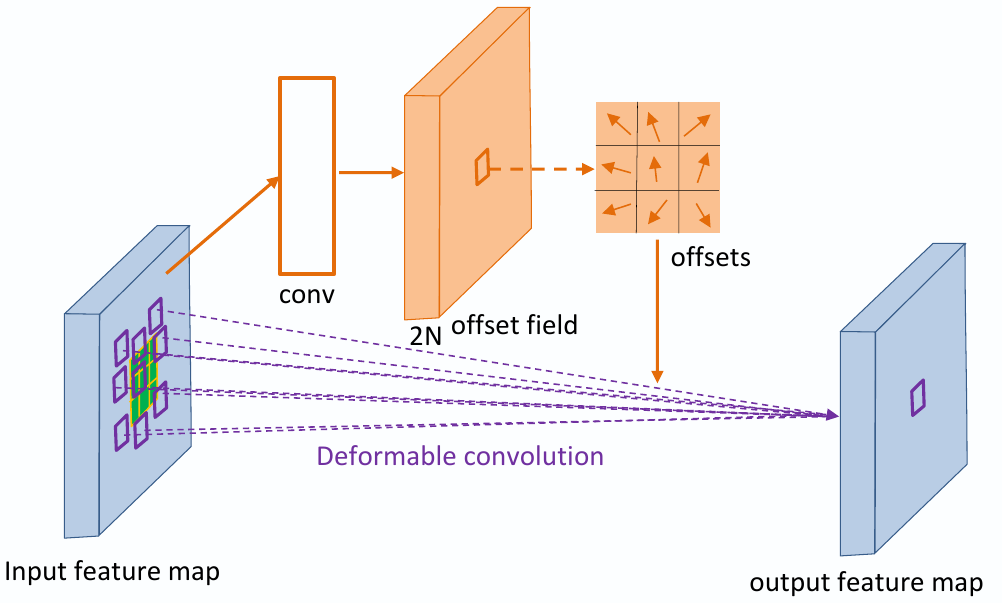}
	\caption{Illustration of $3 \times 3$ deformable convolution. }
	\label{Deforable_conv}
\end{figure}

\begin{figure}
\centering
\subfigure
{\begin{minipage}[t]{0.5\linewidth}
\centering
\includegraphics[scale=0.3]{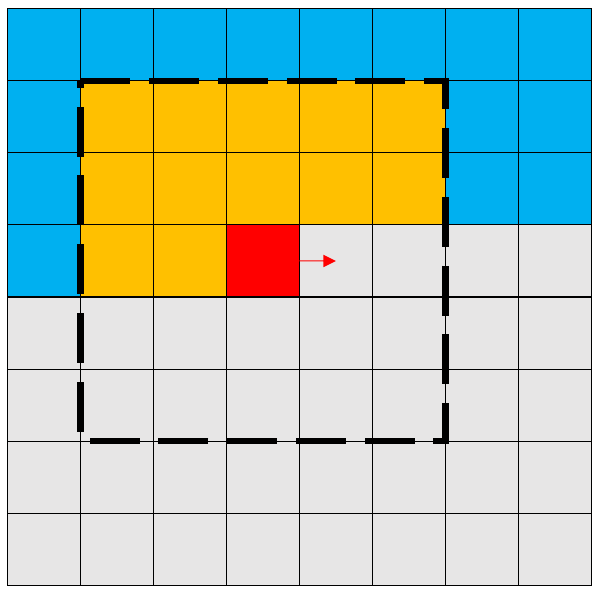}
\label{fig:joint}
\end{minipage}
}%
\subfigure
{\begin{minipage}[t]{0.5\linewidth}
\centering
\includegraphics[scale=0.3]{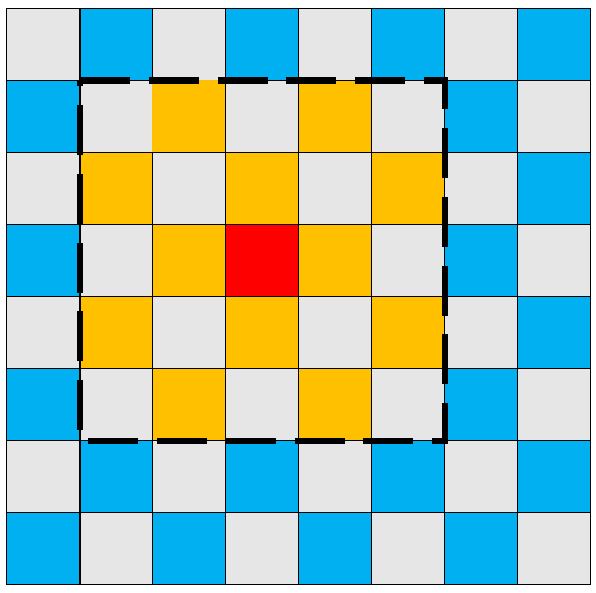}
  \label{checkerboard_mask}
\end{minipage}
}%
\centering
  \caption{ (a) The serial autoregressive context model. Red cell: the symbol to encode/decode. Orange and blue cells: causal neighbors. Orange cells are examples with a $5 \times 5$ convolution window. (b) The checkerboard context model with a $5 \times 5$ window. The first pass decodes all blue and orange anchor cells. The second pass decodes all non-anchor cells.}
  \label{fig:tecnick}
\end{figure}

\begin{figure*}[!thp]
	\centering
		\includegraphics[scale=0.6]{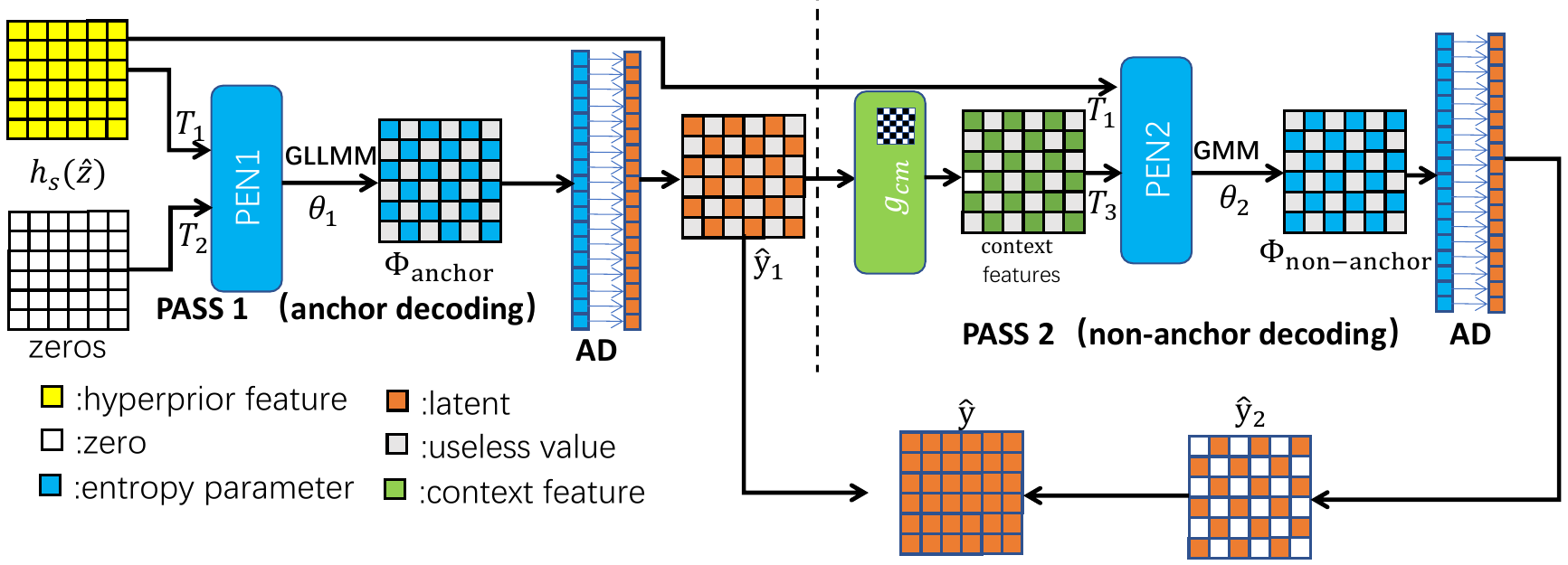}
	\caption{The details of the proposed checkerboard context model scheme.}
	\label{two_pass_decoding}
\end{figure*}

The proposed learned image compression scheme is shown in Fig. \ref{whole_networkstructure}. The input image $x$ has a size of $W\times H\times 3$, where $W$ and $H$ are the width and height of $x$, respectively. The codec mainly includes the core networks ($g_{a}$ and $g_{s}$), the hyper networks ($h_{a}$ and $h_{s}$), and improved checkerboard context model.

The core encoder network $g_{a}$ learns a compact latent representations $y$ of the input image. $g_{a}$ is same as that in \cite{cheng2020}, which includes two simplified attention modules, three residual blocks (shown in gray in Fig. \ref{whole_networkstructure}), and four stages of pooling operators. The difference is that  deformable residual module (DRM) is proposed in this paper.

To enable parallel entropy decoding of the quantized latents $\hat{y}$, it is divided into two checkerboard subsets $\hat{y}_1$ and $\hat{y}_2$. The probability distribution parameters for the two subsets are estimated by two parameter estimation networks (PENs) separately via a two-pass approach. The details are described in Sec. \ref{checkerboard_context} and Fig. \ref{fig:PEN}.

Next, arithmetic coding is used to compress $\hat{y}$ into the bitstream. The decoded $\hat{y}$ is sent to the main decoder $g_{s}$, which is symmetric to the core encoder network $g_{a}$, with convolutions replaced by deconvolutions. The leaky ReLU is used in most convolution layers, except for the last layer in hyperprior encoder and decoder, which does not have any activation function.

Experimental results show that the decoder network's complexity can be lower than the encoder without affecting the reconstruction performance. In this paper, we develop an improved knowledge distillation method to reduce the complexity of the decoder in Fig. \ref{whole_networkstructure}, which serves as a teacher network. The details to obtain the student network from the teacher network are described in Sec. \ref{sec_distillation}.

\subsection{Deformable Residual Block (DRB)}
\label{sec_msrb}

The deformable convolution first was proposed in \cite{Defor_Conv_V1}, and it has been widely used in many fields, including learned video compression. Its structure is shown in Fig. \ref{Deforable_conv}.  
Deformable convolution offers significant benefits by allowing flexible modeling of receptive fields. This helps in extracting better features and representing objects effectively in convolutional neural networks. Consequently, it improves performance in tasks that require precise spatial understanding and object detection. This innovation has the potential to enhance convolutional architectures in capturing complex spatial relationships, making it a promising approach for different computer vision applications.

% The formula for Deformable Convolution is illustrated in Equation

% \begin{equation}
%     y(p_{0}) = \sum_{p_{n} \in R}{\omega(p_{0})\cdot x(p_{0}+p_{n}+\Delta p_{n}) }
%     \label{eq:deform_conv}
% \end{equation}

As depicted in Fig. \ref{Deforable_conv}, the dimensions of the offset field align with those of the input feature map, while 2N corresponds to the channel numbers.

In this paper, we propose a deformable residual module (DRM) and apply it to image compression, as depicted in Fig. \ref{whole_networkstructure}. In our DRM, we combine the deformable module with the classical convolution, and add a shortcut connection. The DRM is used for upsampling or downsampling.  The proposed DRM can be utilized to reduce spatial redundancy in input image, thereby enhancing image compression performance. In the ablation experiment section, we will demonstrate the effectiveness of this module. As in \cite{Defor_Conv_V1}, the deformable module  hardly increases the model complexity compared to the classical convolutions.

\subsection{Improved Checkerboard Context Model and Coding}
\label{checkerboard_context}

Previous learned image compression methods use serial context-adaptive entropy model. Its decoding cannot be parallelized, as shown in Fig. \ref{fig:joint}. To address this issue, a checkerboard context model is proposed in \cite{He_2021_CVPR}, where the latent representation $y$ is divided into two subsets, denoted as anchors $\hat{y}_{1}$ and non-anchors $\hat{y}_{2}$, as shown in Fig. \ref{checkerboard_mask}. The first pass is to encode and decode the anchors. The second pass is to encode and decode the non-anchors based on anchors. Compared to serial context model used in \cite{cheng2020}, the decoding of \cite{He_2021_CVPR} is about $2.5-2.7$ times faster.

However, the R-D performance of \cite{He_2021_CVPR} is dropped by about 0.2-0.3 dB on the Kodak dataset compared to the serial context model used in \cite{cheng2020}. There are two reasons for the drop. First, the anchor part is coded using only hyperprior, but without using any context model. Second, a single network is used to estimate the probability distribution parameters of the two subsets.

In this paper, we propose two techniques to improve the R-D performance of the checkerboard-based approach. First, we use two different networks to estimate the probability distribution parameters of the two subsets separately. Next, since the anchor is coded without context model, it should use more powerful probability distribution model to improve the performance. In this paper, we use the more advanced GLLMM model in \cite{GLLMM} for the anchor part. The non-anchor part still uses the GMM model, as in \cite{He_2021_CVPR}.

The improved checkerboard context model and decoding are shown in Fig. \ref{whole_networkstructure} and Fig. \ref{two_pass_decoding}. During encoding and training, we first obtain the anchors $\hat{y}_{1}$ and non-anchors $\hat{y}_{2}$. Both have the same size as $\hat{y}$. Since the values of the $\hat{y}$ are visible during training and encoding, we just copy the values of $\hat{y}$ to obtain $\hat{y}_{1}$ and $\hat{y}_{2}$. In the first pass, we only encode and train the anchors (blue cells in Fig.\ref{checkerboard_mask} and Fig. \ref{two_pass_decoding}), which only depend on hyperprior and do not adopt any context model. The non-anchors $\hat{y}_{2}$ (grey cells in Fig.\ref{checkerboard_mask} and Fig. \ref{two_pass_decoding}) are coded using both checkerboard context and the hyperprior.

During decoding, since we do not know the values of all latent representations $\hat{y}$, we have to decode the anchors $\hat{y}_{1}$ and non-anchors $\hat{y}_{2}$ in turn, as shown in Fig. \ref{two_pass_decoding}. $\hat{y}_{1}$ and $\hat{y}_{2}$ are initialized to zero tensors, which have the same size as $\hat{y}$. We first utilize the hyper decoder $h_{s}$ to obtain the output $T_{1}$. $T_{1}$ and a zero tensor $T_{2}$ are first combined and sent to network PEN1 to estimate the probability distribution parameters of the anchors, denoted as $\theta_{1}$. Different from \cite{He_2021_CVPR}, we use the more powerful GLLMM model in \cite{GLLMM} to estimate the parameters of the anchors, to improve the performance even when context model is not used. However, the absence of context model enables us to decode all anchors in parallel.

The decoded anchors are then used to update $\hat{y}_{1}$, which will pass though a single convolution layer with checkerboard mask (as shown in Fig. \ref{checkerboard_mask}) to obtain context feature $T_{3}$. $T_{3}$ is then combined with $T_{1}$ and sent to another network PEN2 to estimate the probability distribution parameters of the non-anchors, denoted as $\theta_{2}$. The non-anchors can also be obtained in parallel. Since PEN1 and PEN2 are trained separately, they can achieve better performance than \cite{He_2021_CVPR}, which only uses one network for both anchors and non-anchors. Since context model is already used for non-anchors, the probability model can be simpler. Therefore only GMM model is used for the non-anchors, as in \cite{He_2021_CVPR}. 

The details of the two parameter estimation networks PEN1 and PEN2 are shown in Fig. \ref{fig:PEN}, where as in \cite{GLLMM,cheng2020}, $15N$ and $4.5N$ are the number of parameters of GLLMM and GMM models respectively.

Finally, we can combine $\hat{y}_{1}$ and $\hat{y}_{2}$ to obtain the decoded $\hat{y}$.

\subsection{Improving the Decoder Using Knowledge Distillation}
\label{sec_distillation}

In this part, we use the knowledge distillation to reduce the complexity of the decoder network in Fig. \ref{whole_networkstructure}. In fact, experimental results show that sometimes knowledge distillation can also improve the R-D performance, because the teacher network can transfer some prior knowledge to the student network. Therefore our entire training includes three steps.

\begin{figure*}[!thp]
	\centering
		\includegraphics[scale=0.75]{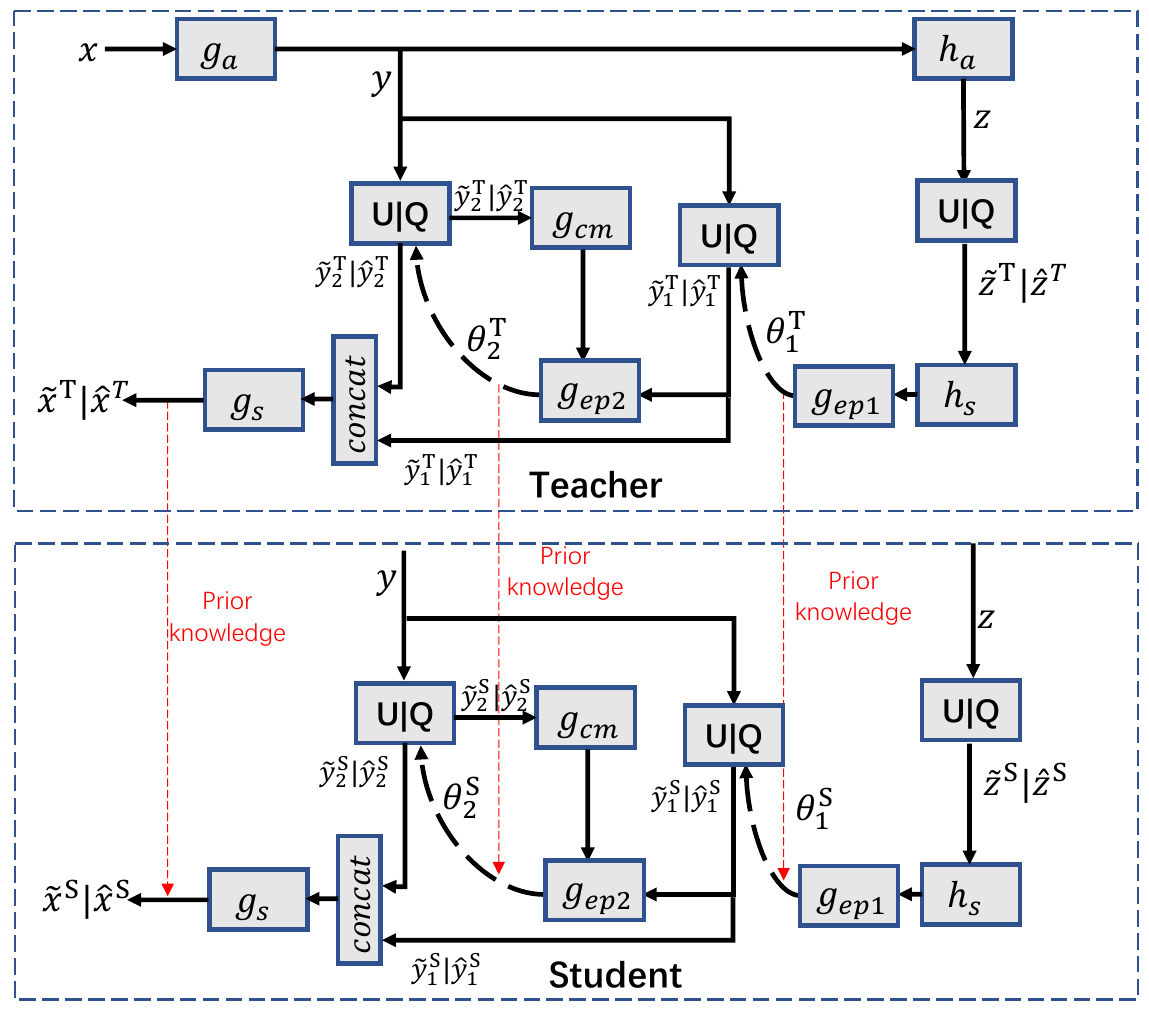}
	\caption{The knowledge distillation framework between the teacher and student decoder networks.}
	\label{distillation_framework_model}
\end{figure*}

First, we train the encoder and the decoder in Fig. \ref{whole_networkstructure} using the following traditional loss function:
\begin{equation}\label{teacher_loss}
\begin{aligned}
 L_{T} =  \lambda_{1} D(x,\hat{x})&+H(\hat{y})+H(\hat{z})+\lambda_{2}L_{1}(\hat{y}), \\
      H(\hat{y}) &=  E [-\log_{2}(P_{\hat{y}|\hat{z}}(\hat{y}|\hat{z}))],\\
      H(\hat{z}) &=  E [-\log_{2}(P_{\hat{z}}(\hat{z}))],
\end{aligned}
\end{equation}
where $D(x, \hat{x})$ is the reconstruction error between the origin image $x$ and the reconstructed image $\hat{x}$. The Mean Squared Error (MSE) and MS-SSIM are considered in this paper. $H(\hat{y})$, $H(\hat{z})$ are the entropies of the core latent representation and hyper representation. $L_{1}$ is  $L_{1}$ norm regularization.

After the training above, we introduce a new student decoder network, which initially has the same architecture as the teacher decoder network in Fig. \ref{whole_networkstructure}. Our goal is to use the knowledge distillation to improve the R-D performance of the student decoder network. In \cite{GAN_distillation_image_compression}, only the prior knowledge of the final reconstruction image is transferred to the student network. In this paper, we also transfer the prior knowledge of the probability distribution parameters $\theta_{1}$ and $\theta_{2}$ to the student network. Our knowledge distillation framework can be illustrated by the block diagram in Fig. \ref{distillation_framework_model}, where superscripts $T$ and $S$ represent teacher and student respectively.

The encoder network, teacher and student decoder networks are jointly trained again, using the following loss function.
\begin{equation}\label{total_loss}
\begin{aligned}
 L_{S} &=  L_{T} +  \lambda_{3} L_{KD}, \\
      L_{KD} &=  d(\hat{x}^{T}, \hat{x}^{S}) + d(\theta_{1}^{T}, \theta_{1}^{S})) + d(\theta_{2}^{T}, \theta_{2}^{S}),
\end{aligned}
\end{equation}
where $L_{T}$ is the loss function in Eq. \ref{teacher_loss}. $L_{KD}$ is the knowledge distillation loss function, which includes the distortions between the teacher and student decoder networks in terms of the reconstructed image, probability distribution parameters $\theta_{1}$ and $\theta_{2}$. Different loss functions can be used in $L_{KD}$. In \cite{hinton2015distilling,GAN_distillation_image_compression}, Softmax is used. In this paper, we find that MSE gives better results, as will be shown in the ablation experiments in Sec. \ref{sec_results}. The prior knowledge is thus transferred from the teacher network to the student network via the loss function $L_{S}$. 

After the joint training above, we can further reduce the complexity of the student network. This is desired in many real-time applications. In this paper, we use the decoder network in \cite{cheng2020} as the baseline, and explore different knowledge distillation techniques to reduce its complexity. For example, we can reduce the number of filters $N$ in the final latent representation, or remove some modules that have higher complexity but do not contribute too much to the performance, such as the attention modules and residual modules.

To optimize the low-complexity student decoder network, we jointly train the encoder, teacher and student decoder networks again using the joint loss function in Eq. \ref{total_loss}. Ablation experiments will be reported in Sec. \ref{sec_results}.

\begin{figure*}
\centering
\subfigure
{\begin{minipage}[t]{0.5\linewidth}
\centering
\includegraphics[width=\columnwidth]{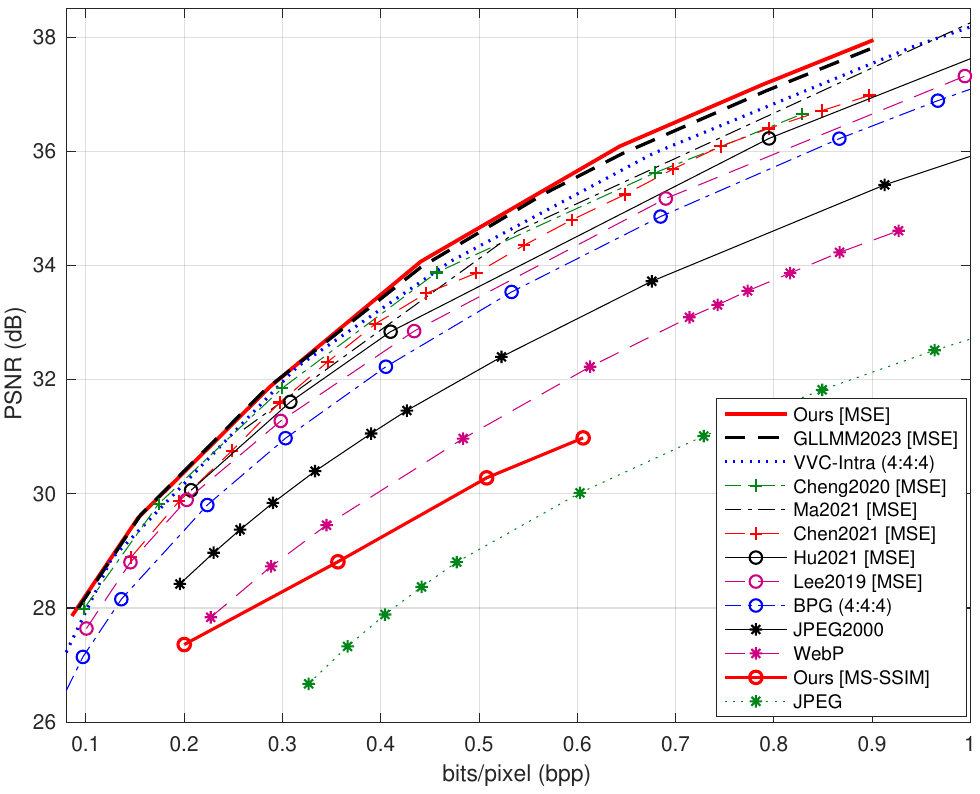}
\label{kodak_PSNR}
\end{minipage}
}%
\subfigure
{\begin{minipage}[t]{0.5\linewidth}
\centering
\includegraphics[width=\columnwidth]{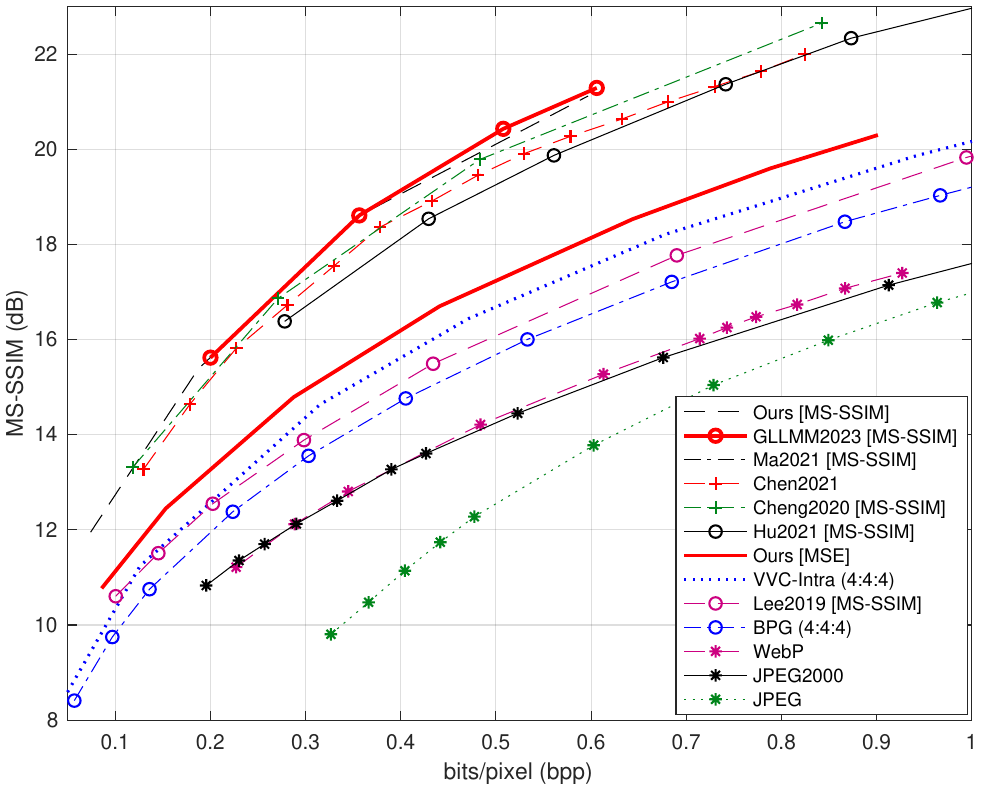}
\label{kodak_SSIM}
\end{minipage}
}%
\centering
\caption{The R-D curves of different methods in terms of PSNR and MS-SSIM on the Kodak dataset \cite{Kodak}.}
\label{fig:kodak}
\end{figure*}

\begin{figure*}
\centering
\subfigure
{\begin{minipage}[t]{0.5\linewidth}
\centering
\includegraphics[width=\columnwidth]{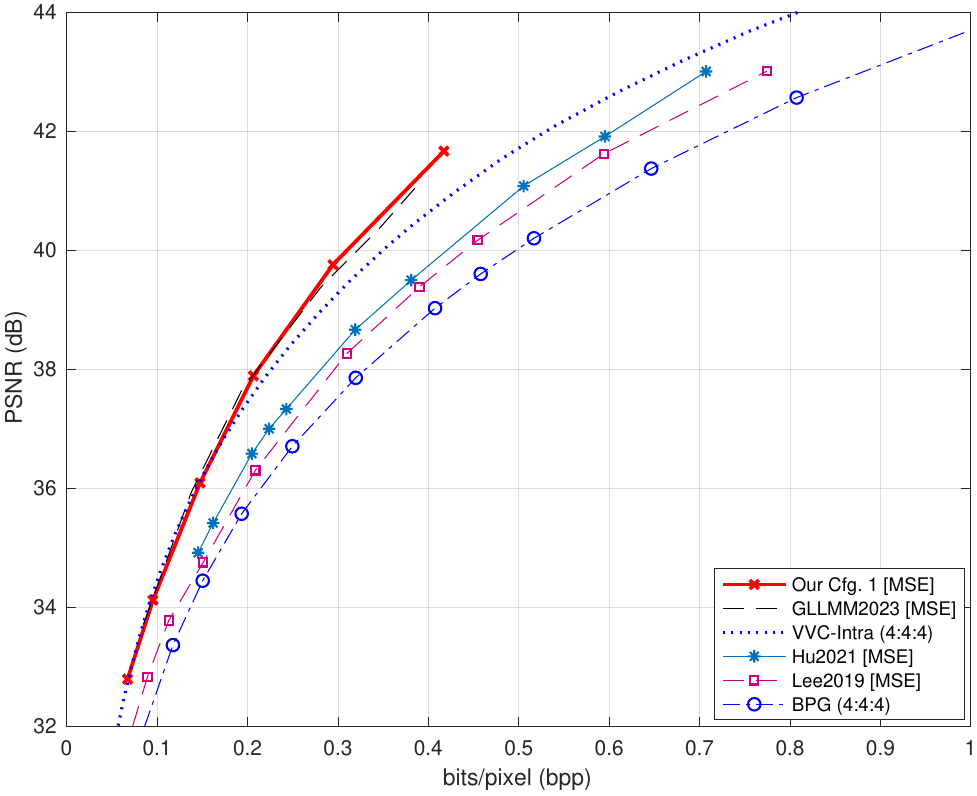}
\label{kodak_PSNR}
\end{minipage}
}%
\subfigure
{\begin{minipage}[t]{0.5\linewidth}
\centering
\includegraphics[width=\columnwidth]{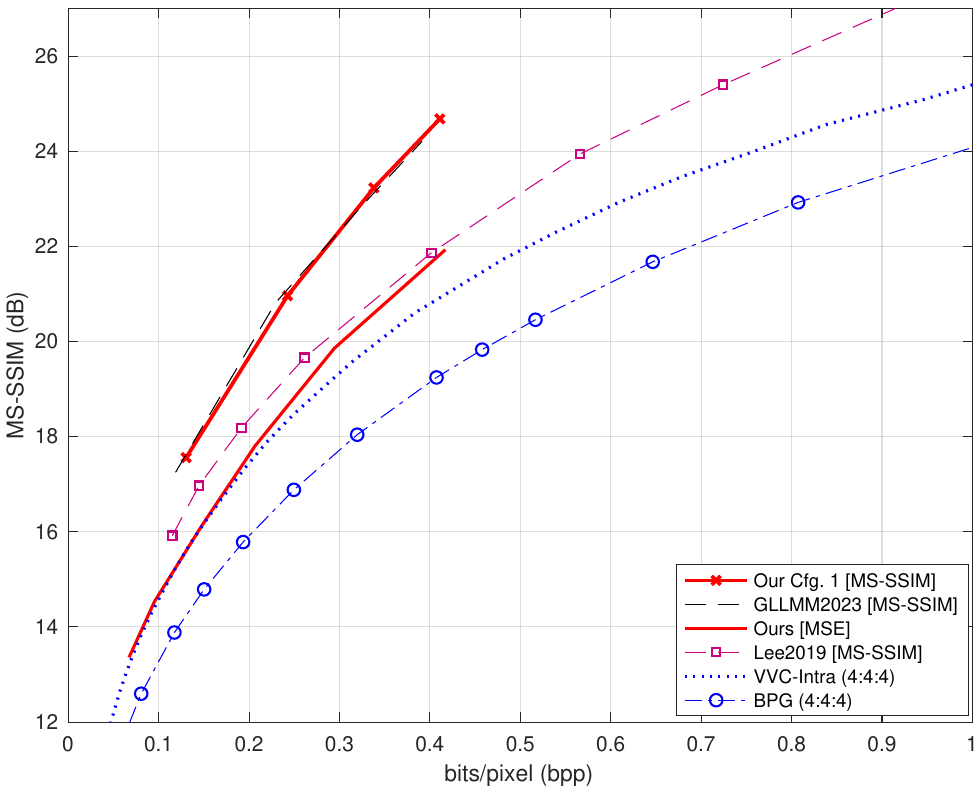}
\label{kodak_SSIM}
\end{minipage}
}%
\centering
\caption{The R-D curves of different methods in terms of PSNR and MS-SSIM on the Tecnick-40 dataset \cite{Tecnick}.}
\label{fig:tecnick}
\end{figure*}

% \begin{figure*}
% \centering
% \subfigure
% {\begin{minipage}[t]{0.33\linewidth}
% \centering
% \includegraphics[scale=0.4]{images/examples/keda_15_origin_compare.png}
% \caption{Original}
% \end{minipage}
% }%
% \subfigure
% {\begin{minipage}[t]{0.33\linewidth}
% \centering
% \includegraphics[scale=0.4]{images/examples/keda_15_VVC_compare.png}
% \caption{VVC (0.108/30.62/0.945)}
% \end{minipage}
% }%
% \subfigure
% {\begin{minipage}[t]{0.33\linewidth}
% \centering
% \includegraphics[scale=0.4]{images/examples/keda_15_ours_compare.png}
% \caption{Ours (0.101/30.68/0.9469)}
% \end{minipage}
% }%
% \caption{Visual examples of different image compression methods. Our method is optimized for PSNR. The numbers reported are bit rate (BPP), PSNR (dB), and MS-SSIM.}
%  \label{fig:Example}
% \end{figure*}

\subsection{Training}

The training images are collected from the CLIC dataset \cite{CLIC} and LIU4K dataset \cite{LIU_dataset}. All training images are rescaled to a resolution of $2000 \times 2000$. We also utilize some data augmentation technologies such as rotation and scaling to obtain 81,650 training images with a resolution of $384 \times 384$.

Both mean squared error (MSE) and multi-scale structural similarity (MS-SSIM) are considered as distortion to optimize our models. For MSE optimization, $\lambda_{1}$ is chosen from the set $\{0.0016,0.0032,0.0075,0.015, 0.03, 0.045, 0.06\}$. Each $\lambda_{1}$ trains an independent model for a bit rate. The number of filters $N$ in the latent representation is set to 128 for the first three $\lambda_{1}$, and is increased to 256 for the last four cases. For MS-SSIM optimization, $\lambda$ is set to 12, 40, 80, and 120 respectively. The value of $N$ is set to 128 for $\lambda$ = 40 and 80, and 256 for $\lambda$ = 80 and 120. Each model is trained for $1.5 \times 10^{6}$ iterations. The Adam solver with a batch size of 8 is adopted. The learning rate is set to $1 \times 10^{-4}$ in the first 750,000 iterations. After that, it is reduced by 0.5 after every 100,000 iterations. $\lambda_{2}$ is set to 0.0001 in the first 10,000 iterations, and is set to 0 after 10,000 iterations. $\lambda_{3}$ is set to 1 in the first 20,000 iterations, and is set to 0 after 20,000 iterations. That is, the knowledge distillation is used at the beginning to pass the prior knowledge to the student network. After that, there is no need to have the $L_{KD}$ term in Eq. \ref{total_loss} to reduce the training complexity.

\section{Experimental Results}
\label{sec_results}

In this section, we compare the proposed method with some state-of-the-art learning-based image compression approaches and traditional methods in terms of PSNR and MS-SSIM using both Kodak PhotoCD dataset \cite{Kodak} and Tecnick-40 dataset \cite{Tecnick}. The learned image compression methods include GLLMM \cite{GLLMM}, He2021 \cite{He_2021_CVPR}, Hu2020 \cite{Hu_AAAI}, Cheng2020 \cite{Cheng2020_paper}, and Lee2019 \cite{Lee_2020}. The classical methods include the latest VVC-Intra (4:4:4) \cite{VVC}, BPG-Intra (4:4:4),  JPEG2000, and JPEG. The Kodak dataset has 24 test images with a resolution at $768 \times 512$, The Tecnick-40 dataset has 40 test images with a size of $1200 \times 1200$. The PSNR and MS-SSIM are chosen as evaluation metrics.

We present our results with four optimized decoder configurations. Cfg. 1 has the same decoder architecture as in Fig. \ref{whole_networkstructure}. Based on Cfg. 1, Cfg. 2 only removes the attention and residual modules, Cfg. 3 only reduces all $N$ by $25 \%$, and Cfg. 4 only reduces all $N$ by $50 \%$.

Note that for fair comparison, we implement the method in Cheng2020 \cite{Cheng2020_paper} and increase its number of filters $N$ from 192 to 256 at high rates, which leads to better performance than the original results in \cite{Cheng2020_paper}. The results of He2021 \cite{He_2021_CVPR} are based on the source code at \cite{lu2022high}.

\subsection{R-D Performances}

\begin{figure*}[!tph]
\centering
\subfigure[Original]{
\begin{minipage}[t]{0.33\linewidth}
\centering
\includegraphics[scale=0.31]{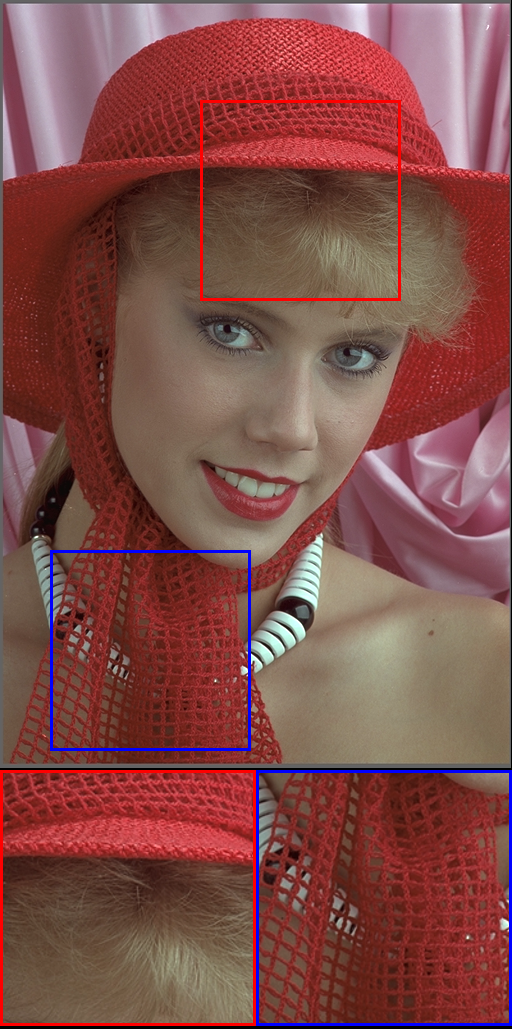}
\end{minipage}
}%
\subfigure[JPEG (0.156/21.28/0.651)]{
\begin{minipage}[t]{0.33\linewidth}
\centering
\includegraphics[scale=0.31]{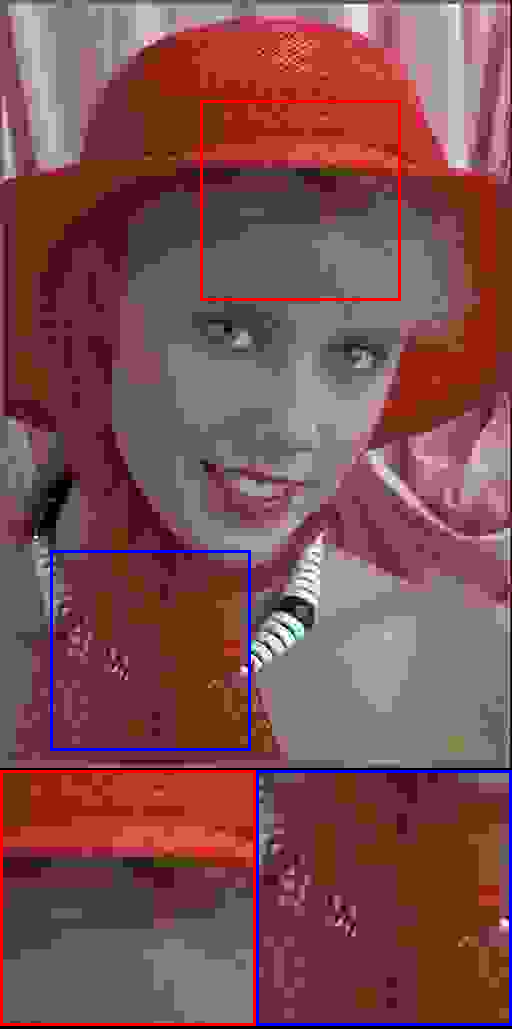}
\end{minipage}
}%
\subfigure[JPEG2000(0.116/29.22/0.904)]{
\begin{minipage}[t]{0.33\linewidth}
\centering
\includegraphics[scale=0.31]{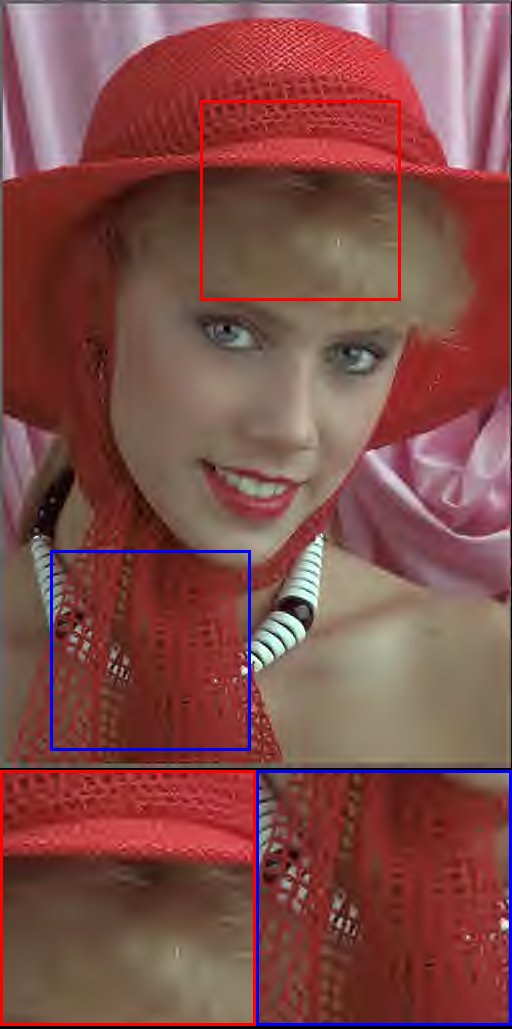}
\end{minipage}
}%

\subfigure[BPG(0.106/30.02/0.916]{
\begin{minipage}[t]{0.33\linewidth}
\centering
\includegraphics[scale=0.31]{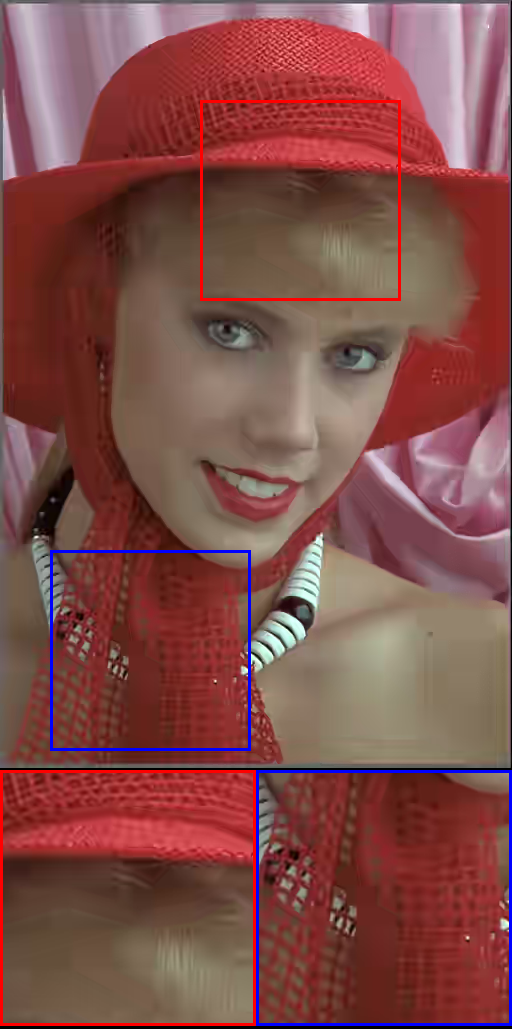}
\end{minipage}
}%
\subfigure[VVC(0.103/30.90/0.929)]{
\begin{minipage}[t]{0.33\linewidth}
\centering
\includegraphics[scale=0.31]{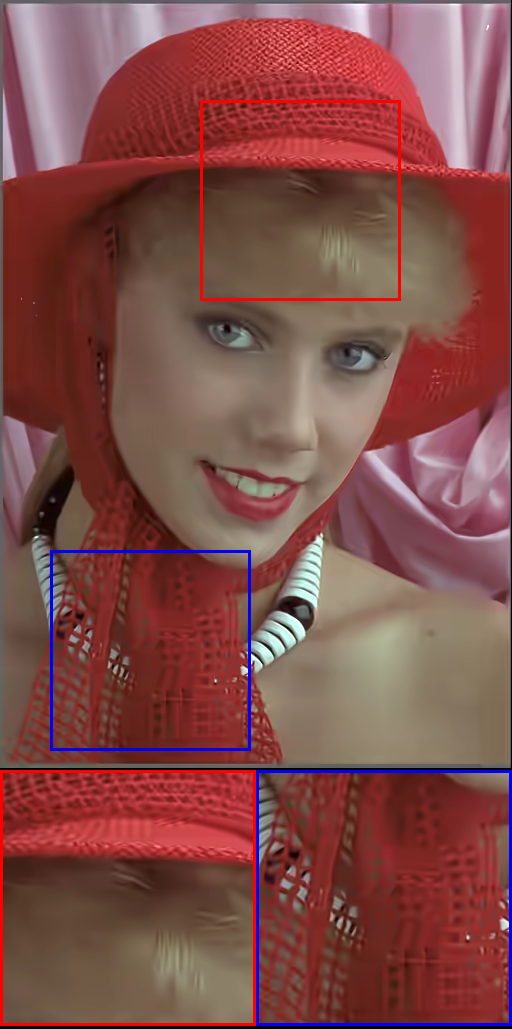}
\end{minipage}
}%
\subfigure[Ours(0.101/31.05/0.932)]{
\begin{minipage}[t]{0.33\linewidth}
\centering
\includegraphics[scale=0.31]{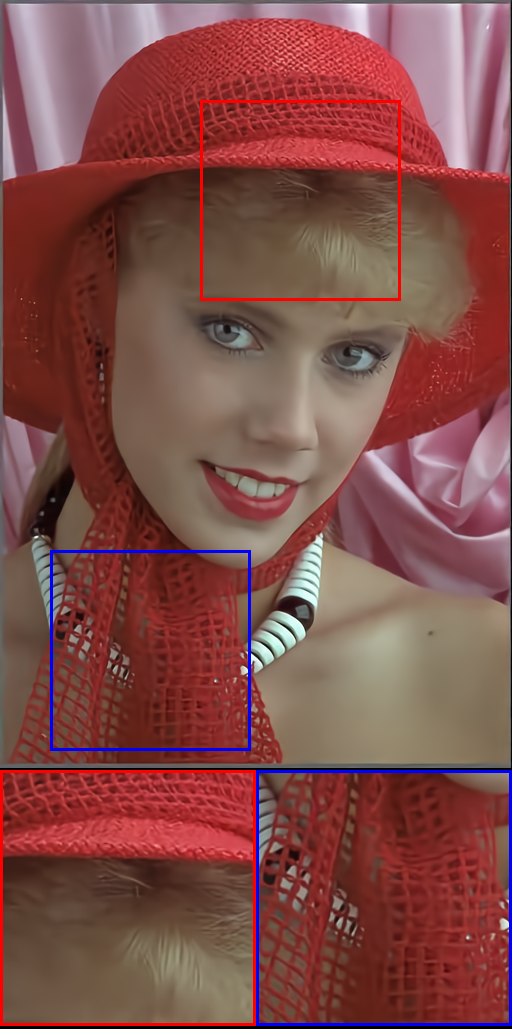}
\end{minipage}
}%
\centering
\caption{Visual examples of different image compression methods. Our method is optimized for PSNR. The numbers reported are bit rate (BPP), PSNR (dB), and MS-SSIM.}
\label{fig:Example}
\end{figure*}

The average R-D curves of different methods on the Kodak dataset are shown in Fig. \ref{fig:kodak}. When optimized for PSNR, GLLMM (MSE) \cite{GLLMM} obtains the best performance among the competing methods, which also outperforms VVC (4:4:4). Our Cfg. 1 achieves the same performance with GLLMM at low bit rates and has better performance at high bit rates.  Our Cfg. 1 achieves the same performance with VVC (4:4:4) at low bit rates. When the bit rate is higher than 0.4 bpp, our Cfg. 1 has a gain of 0.25-0.3 dB over VVC (4:4:4). When optimized for MS-SSIM, our method is also slightly higher than GLLMM. A visual example is given in Fig. \ref{fig:Example}.

Fig. \ref{fig:tecnick} shows the results on the Tecnick-40 dataset. When the bit rate is lower than 0.2 bpp, Our Cfg. 1 achieves the same performance with GLLMM. When the bit rate is higher than 0.2 bpp,  Our Cfg. 1 is sightly better than GLLMM \cite{GLLMM}. Our Cfg. 1 also outperforms other learning-based methods and traditional image codecs.

\subsection{Complexity and Performance Trade-off}

\begin{table*}[!t]
\caption{Comparisons of encoding and decoding time, BD-Rate saving over VVC, and model sizes.}
\begin{center}
\begin{tabular}{|c|c|c|c|c|c|c|}
\hline
\textbf{Dataset}& \textbf{Method} & \textbf{Encoding time} & \textbf{Decoding time} & \textbf{BD-Rate} &\textbf{Model size(Low) }&\textbf{Model size(High)} \\ 
\hline
\multirow{10}{*}{Kodak}  & VVC  & 402.27s& 0.607s& 0.0 & 7.2 MB & 7.2MB\\ \cline{2-7} %%%FIRST BIT RATE
                         & Lee2019 \cite{Lee_2020}  &10.721s& 37.88s& 17.0\% & 123.8 MB & 292.6MB\\ \cline{2-7} %%%FIRST BIT RATE
                         & Hu2021 \cite{Hu_2021}  &35.7187s& 77.3326s& 11.1 \% & 84.6 MB & 290.9MB\\ \cline{2-7} %%%FIRST BIT RATE
                         & Cheng2020 \cite{cheng2020}  &26.37s& 28.46s& 2.6 \% & 50.8 MB & 175.18MB\\
                         \cline{2-7} %%%FIRST BIT RATE
                         & He2021 \cite{He_2021_CVPR}  &24.36s& 5.21s& 8.9 \% & 46.6 MB & 156.6 MB\\
                         \cline{2-7} %%%FIRST BIT RATE
                         & GLLMM \cite{GLLMM}  &467.90s& 467.90s& -3.13\% & 77.08 MB & 241.03MB\\ \cline{2-7} %%%FIRST BIT RATE
                         & \textbf{Our Cfg. 1}  &\textbf{25.08 s}& \textbf{4.45s}& \textbf{-4.25\%} & \textbf{63.06 MB} &  \textbf{197.8MB}\\ \cline{2-7}
                          & \textbf{Our Cfg. 2}  &\textbf{24.02 s}& \textbf{3.03s}& \textbf{-1.89\%} & \textbf{54.26 MB} &  \textbf{166.9MB}\\\cline{2-7} 
                          & \textbf{Our Cfg. 3}  &\textbf{22.56 s}& \textbf{2.78s}& \textbf{-0.19\%} & \textbf{54.66 MB} &  \textbf{164.1MB}\\\cline{2-7} 
                           & \textbf{Our Cfg. 4}  &\textbf{18.24 s}& \textbf{2.45s}& \textbf{14.23\%} & \textbf{47.6 MB} &  \textbf{134.1MB}\\ \cline{2-7}
                         \hline%%%FIRST BIT RATE
                         
\multirow{10}{*}{Tecnick}  & VVC  & 700.59s& 1.49s& 0.0 & 7.2 MB & 7.2MB\\ \cline{2-7} %%%FIRST BIT RATE
                         & Lee2019 \cite{Lee_2020} &54.8s& 138.81s& 31.59 \%  & 123.8 MB & 292.6MB\\ \cline{2-7} %%%FIRST BIT RATE
                         & Hu2021 \cite{Hu_2021} &84.035s& 271.50s& 23.06 \% & 84.6 MB & 290.9MB\\ \cline{2-7} %%%FIRST BIT RATE
                         & Cheng2020 \cite{cheng2020} &59.48s& 71.71s& 5.93 \% & 50.8MB &175.18MB\\
                         \cline{2-7} %%%FIRST BIT RATE
                         &He2021 \cite{He_2021_CVPR}  &56.26s& 12.45s& 12.21 \% & 46.6 MB & 156.6 MB\\
                         \cline{2-7} %%%FIRST BIT RATE
                         & GLLMM \cite{GLLMM}   &1233.05s& 1245.05s& -5.14\%& 77.08 MB & 241.03MB\\ \cline{2-7} %%%FIRST BIT RATE
                         &\textbf{Our Cfg. 1}  &\textbf{57.63}s& \textbf{11.65s}& \textbf{-5.27\%} & \textbf{63.06 MB} & \textbf{197.8 MB}\\ \cline{2-7}
                         &\textbf{Our Cfg. 2}  &\textbf{50.47}s& \textbf{7.65s}& \textbf{-2.38\%} & \textbf{54.26  MB} & \textbf{166.9 MB}\\ \cline{2-7}
                         &\textbf{Our Cfg. 3}  &\textbf{46.56}s& \textbf{5.23s}& \textbf{-1.20\%} & \textbf{54.66  MB} & \textbf{164.1 MB}\\ \cline{2-7}
                           &\textbf{Our Cfg. 4}  &\textbf{38.67}s& \textbf{4.78 s}& \textbf{15.78\%} & \textbf{47.6 MB} & \textbf{134.1 MB}\\ \cline{2-7}
                         \hline%%%FIRST BIT RATE
\end{tabular}

\label{runing_time}
\end{center}
\end{table*}

Table \ref{runing_time} compares the average encoding/decoding time, BD-Rate saving over VVC \cite{BDRate}, and model sizes at low rate and high rate for different methods. Since VVC, Hu2020 \cite{Hu_AAAI}, and Cheng2020 \cite{cheng2020} suffer from a non-determinism issue \cite{Sun_2021} on GPU and only run on CPU, we test on an 2.9GHz Intel Xeon Gold 6226R CPU.

Compared to the state-of-the-art GLLMM method \cite{GLLMM}, our Cfg. 1 is about 20 times faster in encoding and 70-90 times faster in decoding, and our R-D performance is better. Our model size is also smaller.

Compared to Cheng2020 \cite{cheng2020}, our Cfg. 1 encoding time is similar, but decoder is about 4-5 times faster. Our R-D performance is $6.85\%$ and $11.20\%$ better. Our speed is similar to \cite{He_2021_CVPR}, but our R-D performance is about $15 \%$  better.

Our Cfg. 2 and Cfg. 3 can further reduce the decoder complexity by  $20-30 \%$, with $2.6-4.0\%$ loss in R-D performance compared to Cfg. 1, but still have better performance than other learning-based methods and VVC (4:4:4). Cfg. 4 is faster but has $18.3 \%$ drop in R-D performance. Therefore our method can offer various trade-offs between complexity and R-D performance.

\subsection{Ablation Experiments}

\begin{table*}[!thp]
\begin{center}
  \begin{tabular}{ccccccc}
  \hline
 \textbf{Name} & \textbf{Bit rates} & \textbf{All-Zero Channels} & \textbf{Total Channels}  & \textbf{Dec. Time (Ours)} & \textbf{Dec. Time (Full)} &\textbf{ Dec. Reduction}\\
  \hline
  \textbf{Kodak} & Low &76 & 128    &4.35 s & 6.45s & 48.27\%\\
  \textbf{Kodak} & High &124 & 256  &64.43 s  &100.22 s   & 55.54\%\\
  \hline
  \textbf{Tecnick} & Low &78 & 128   & 11.65 s  & 17.44 s & 49.37 \%\\
  \textbf{Tecnick} & High &123 & 256  & 203.49 s  & 324.58 s &59.50\%\\
  \hline
\end{tabular}
\label{Channel_algorithem}
\end{center}
\caption{The comparison of different decoding methods.}
\end{table*}

 \begin{figure}[!thp]
	\centering
		\includegraphics[scale=0.55]{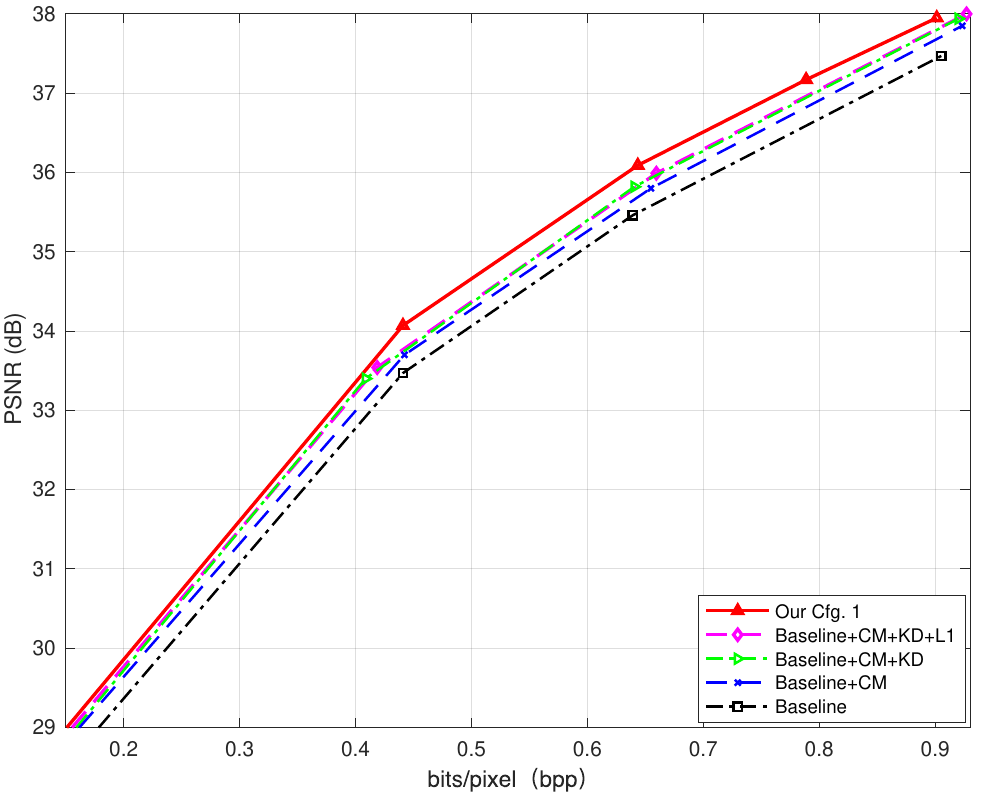}
	\caption{ The contributions of the improved checkerboard context model and the knowledge distillation.}
	\label{Ablation}
\end{figure}

In this part, we show some ablation experiments. All results are the average of the Kodak dataset.

We first show the contributions of the improved checkerboard context model, the knowledge distillation, $L_{1}$ regularization, and deformable residual module in Fig. \ref{Ablation}. We replace the GMM model in \cite{cheng2020} with the checkerboard entropy model \cite{He_2021_CVPR}, and other parts remain unchanged. The modified scheme is used as the baseline.  On top of the baseline, we add different modules in turn.

The results are shown in Fig. \ref{Ablation}.  We first replace the  checkerboard entropy model \cite{He_2021_CVPR} in the baseline by the proposed checkerboard entropy model \cite{He_2021_CVPR}, denoted as Baseline+CM, which improves the R-D  performance by about 0.2-0.3 dB at the same bit rate. Next, we add the knowledge distillation to Baseline+CM, denoted as Baseline+CM+KD. Compared to Baseline+CM, Baseline+CM+KD improves the R-D performance by about 0.1-0.15 dB at the same bit rate. Then, we add the $L_{1}$ regularization to the loss function, denoted as Baseline+CM+KD+L1. It can be observed that introducing $L_{1}$ regularization does not reduce the encoding performance. It just makes the value of latent representation more sparse. Last, we add the deformable residual module to Baseline+CM+KD+L1, which is our proposed method. Compared to Baseline+CM+KD+L1,  the proposed full method will improve the performance  by another 0.1-0.15 dB.

 \begin{figure}[!thp]
	\centering
		\includegraphics[scale=0.55]{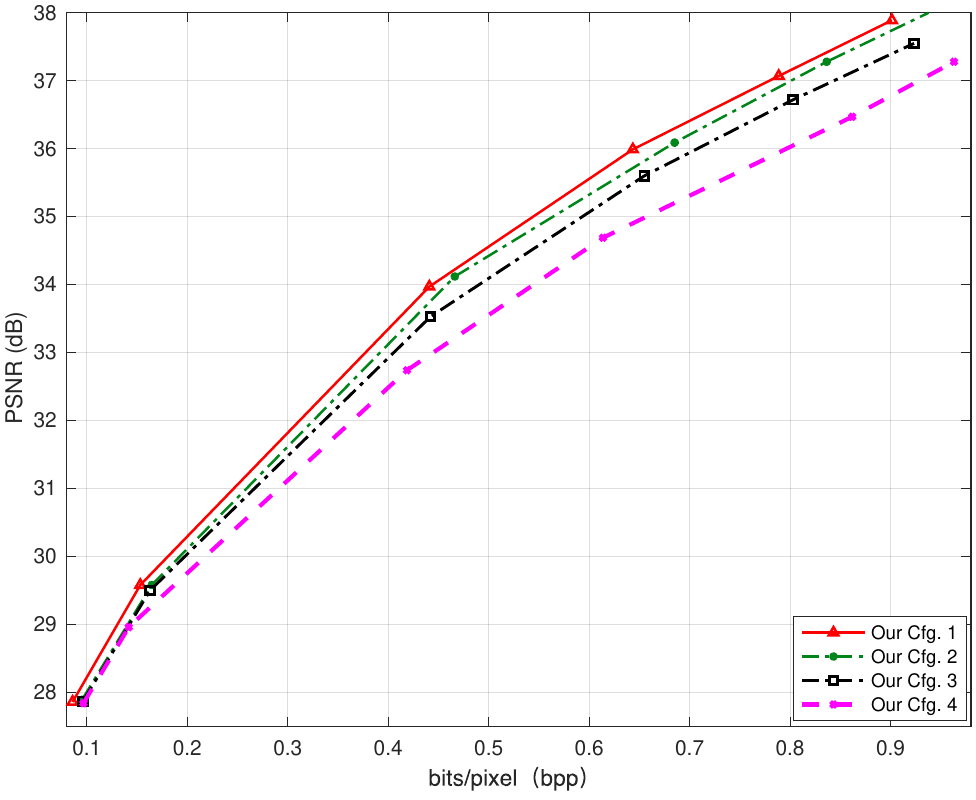}
	\caption{R-D performances of different configurations of the proposed method for Kodak dataset.}
	\label{decoder_complexity}
\end{figure}

Fig. \ref{decoder_complexity} shows the detailed R-D curves of the four configurations of our method. Together with Table \ref{runing_time}, it can be observed that the R-D performances of Cfg. 2 and Cfg. 3 are only slightly lower than Cfg. 1, and the model size is reduced by about $15\%$. The PSNR of Cfg. 4 is more than 1 dB lower than Cfg. 1 at high rates, which shows that at high rates, the network needs more filters to ensure good performance. These results suggest that we can combine different knowledge distillation methods. For example, at low bit rates, we can reduce the number filters. At high bit rates, we can remove the attention models and residual blocks.

\begin{table}[!thp]
\begin{center}
  \begin{tabular}{cccc}
  \hline
 \textbf{Module} & \textbf{Bit rate}  & \textbf{PSNR (dB)} & \textbf{MS-SSIM (dB)}\\
  \hline
  \textbf{Softmax}               & 0.1643 & 29.67 & 12.60 \\
  \textbf{MSE}                    & \textbf{0.1628} &\textbf{29.76}  &\textbf{12.62}\\
  \hline
  \textbf{Softmax} & 0.8046  &37.05 &19.58  \\
 \textbf{MSE} & \textbf{0.8028} &\textbf{37.23} & \textbf{19.68}  \\
  \hline
\end{tabular}
\end{center}
\caption{Comparison of different knowledge distillation losses.}
\label{table_KD_loss}
\end{table}

 Table \ref{table_KD_loss} compares the performance when the Softmax and MSE are used in the knowledge distillation loss function $L_{KD}$, which shows that MSE has better performance.

We introduce the $L_{1}$ regularization to make the latent representation more sparse. It can produce more zeros, and is more likely to skip all-zero channels. Table \ref{Channel_algorithem} shows the number of all-zero channels, total channels, our decoding time when all-zero channels are skipped, decoding time when all channels are coded, and the reduction rate of our decoding time with skipped all-zero channels. It can be observed that our method can save $48-59\%$ decoding time.  We provided an example from the Kodak dataset, as shown in Fig. \ref{fig:L1norm}. It can be observed that the introduction of the L1 norm into the loss function results in a sparser latent representation.

\section{Conclusions}

In this paper, we propose four techniques to improve the R-D performance, speed up the decoding of the learned image compression and reduce its decoder complexity, based on deformable residual module, improved checkerboard context model, knowledge distillation, and $L_{1}$ regularization respectively. We are the first to propose deformable residual module (DRM) to further reduce the spatial redundancy of latent representations and improve the R-D performance. In the checkerboard context model, we use two separate networks to estimate the probability distribution parameters of the two subsets, and we also employ the GLLMM model for the first subset, to compensate its loss of performance since it is not coded using context model. We also develop a three-step knowledge distillation scheme for the decoder and the corresponding training strategy to achieve different trade-offs between complexity and performance.  We also introduce $L_{1}$ regularization to make the numerical values of the latent representation more sparse. Then we only encode non-zero channels in the encoding and decoding process, which can greatly reduce the encoding and decoding time and sacrificing coding performance. 

Experimental results using the Kodak and Tecnick-40 datasets show that our proposed methods not only achieve better performance than the state-of-the-art learning-based image compression methods, but also is 70-90 times faster. It also has better performance than traditional image codecs including the H.266/VVC in both PSNR and MS-SSIM metrics. 

The checkerboard context model and knowledge distillation proposed in this paper can be further optimized in the future.

%\clearpage
\ifCLASSOPTIONcaptionsoff
  \newpage
\fi

\bibliographystyle{IEEEtran}
\bibliography{egbib}{}

\begin{thebibliography}{10}
\providecommand{\url}[1]{#1}
\csname url@samestyle\endcsname
\providecommand{\newblock}{\relax}
\providecommand{\bibinfo}[2]{#2}
\providecommand{\BIBentrySTDinterwordspacing}{\spaceskip=0pt\relax}
\providecommand{\BIBentryALTinterwordstretchfactor}{4}
\providecommand{\BIBentryALTinterwordspacing}{\spaceskip=\fontdimen2\font plus
\BIBentryALTinterwordstretchfactor\fontdimen3\font minus
  \fontdimen4\font\relax}
\providecommand{\BIBforeignlanguage}[2]{{%
\expandafter\ifx\csname l@#1\endcsname\relax
\typeout{** WARNING: IEEEtran.bst: No hyphenation pattern has been}%
\typeout{** loaded for the language `#1'. Using the pattern for}%
\typeout{** the default language instead.}%
\else
\language=\csname l@#1\endcsname
\fi
#2}}
\providecommand{\BIBdecl}{\relax}
\BIBdecl

\bibitem{JPEG}
G.~K. Wallace, ``The jpeg still picture compression standard,'' \emph{IEEE
  Transactions on Consumer Electronics}, vol.~38, no.~1, pp. 18--34, 1992.

\bibitem{JPEG2000}
A.~Skodras, C.~Christopoulos, and T.~Ebrahimi, ``The jpeg 2000 still image
  compression standard,'' \emph{IEEE Signal Processing Magazine}, vol.~18,
  no.~5, pp. 36--58, 2001.

\bibitem{BPG}
G.~J. Sullivan, J.-R. Ohm, W.-J. Han, and T.~Wiegand, ``Overview of the high
  efficiency video coding (hevc) standard,'' \emph{IEEE Transactions on
  Circuits and Systems for Video Technology}, vol.~22, no.~12, pp. 1649--1668,
  2012.

\bibitem{VVC}
\BIBentryALTinterwordspacing
H.~Fraunhofer, ``Vvc official test model vtm,'' 2019. [Online]. Available:
  \url{https://vcgit.hhi.fraunhofer.de/jvet/VVCSoftware_VTM/tree/VTM-5.2}
\BIBentrySTDinterwordspacing

\bibitem{GLLMM}
H.~Fu, F.~Liang, J.~Lin, B.~Li, M.~Akbari, J.~Liang, G.~Zhang, D.~Liu, C.~Tu,
  and J.~Han, ``Learned image compression with gaussian-laplacian-logistic
  mixture model and concatenated residual modules,'' \emph{IEEE Transactions on
  Image Processing}, vol.~32, pp. 2063--2076, 2023.

\bibitem{Chen_TNNLS_2022}
H.~Chen, X.~He, H.~Yang, L.~Qing, and Q.~Teng, ``A feature-enriched deep
  convolutional neural network for jpeg image compression artifacts reduction
  and its applications,'' \emph{IEEE Transactions on Neural Networks and
  Learning Systems}, vol.~33, no.~1, pp. 430--444, 2022.

\bibitem{FU_2020}
H.~Fu, F.~Liang, B.~Lei, N.~Bian, Q.~Zhang, M.~Akbari, J.~Liang, and C.~Tu,
  ``Improved hybrid layered image compression using deep learning and
  traditional codecs,'' \emph{Signal Processing: Image Communication}, vol.~82,
  p. 115774, 2020.

\bibitem{Asymmetric_Fu}
H.~Fu, F.~Liang, J.~Liang, B.~Li, G.~Zhang, and J.~Han, ``Asymmetric learned
  image compression with multi-scale residual block, importance scaling, and
  post-quantization filtering,'' \emph{IEEE Transactions on Circuits and
  Systems for Video Technology}, vol.~33, no.~8, pp. 4309--4321, 2023.

\bibitem{chen2021}
T.~Chen, H.~Liu, Z.~Ma, Q.~Shen, X.~Cao, and Y.~Wang, ``End-to-end learnt image
  compression via non-local attention optimization and improved context
  modeling,'' \emph{IEEE Transactions on Image Processing}, vol.~30, pp.
  3179--3191, 2021.

\bibitem{Li_Nonlocal_entropy}
M.~Li, K.~Zhang, J.~Li, W.~Zuo, R.~Timofte, and D.~Zhang, ``Learning
  context-based nonlocal entropy modeling for image compression,'' \emph{IEEE
  Transactions on Neural Networks and Learning Systems}, vol.~34, no.~3, pp.
  1132--1145, 2023.

\bibitem{xie2021enhanced}
Y.~Xie, K.~L. Cheng, and Q.~Chen, ``Enhanced invertible encoding for learned
  image compression,'' in \emph{Proceedings of the ACM International Conference
  on Multimedia}, 2021, pp. 162--170.

\bibitem{zhu2022transformerbased}
Y.~Zhu, Y.~Yang, and T.~Cohen, ``Transformer-based transform coding,'' in
  \emph{International Conference on Learning Representations}, 2022.

\bibitem{Zou_2022_CVPR}
R.~Zou, C.~Song, and Z.~Zhang, ``The devil is in the details: Window-based
  attention for image compression,'' in \emph{Proceedings of the IEEE/CVF
  Conference on Computer Vision and Pattern Recognition (CVPR)}, June 2022, pp.
  17\,492--17\,501.

\bibitem{Variational}
J.~Ball\'{e}, D.~Minnen, S.~Singh, S.~J. Hwang, and N.~Johnston, ``Variational
  image compression with a scale hyperprior,'' in \emph{International
  Conference on Learning Representations}, 2018, pp. 1--23.

\bibitem{Joint}
D.~Minnen, J.~Ball\'{e}, and G.~D. Toderici, ``Joint autoregressive and
  hierarchical priors for learned image compression,'' in \emph{Advances in
  Neural Information Processing Systems}, 2018, pp. 10\,794--10\,803.

\bibitem{cheng2020}
Z.~Cheng, H.~Sun, M.~Takeuchi, and J.~Katto, ``Learned image compression with
  discretized gaussian mixture likelihoods and attention modules,'' in
  \emph{Proceedings of the IEEE/CVF Conference on Computer Vision and Pattern
  Recognition (CVPR)}, 2020, pp. 7939--7948.

\bibitem{Lee_2021}
J.~Lee, S.~Cho, and M.~Kim, ``Joint autoregressive and hierarchical priors for
  learned image compression,'' \emph{arXiv:1912.12817}, 2020.

\bibitem{Defor_Conv_V1}
J.~Dai, H.~Qi, Y.~Xiong, Y.~Li, G.~Zhang, H.~Hu, and Y.~Wei, ``Deformable
  convolutional networks,'' in \emph{2017 IEEE International Conference on
  Computer Vision (ICCV)}, 2017, pp. 764--773.

\bibitem{resblock}
K.~He, X.~Zhang, S.~Ren, and J.~Sun, ``Deep residual learning for image
  recognition,'' in \emph{Proceedings of the IEEE Conference on Computer Vision
  and Pattern Recognition (CVPR)}, June 2016, pp. 770--778.

\bibitem{hinton2015distilling}
G.~Hinton, O.~Vinyals, and J.~Dean, ``Distilling the knowledge in a neural
  network,'' 2015.

\bibitem{end_to_end}
J.~Ball{\'e}, V.~Laparra, and E.~P. Simoncelli, ``End-to-end optimized image
  compression,'' in \emph{International Conference on Learning
  Representations}, 2017.

\bibitem{channel}
D.~Minnen and S.~Singh, ``Channel-wise autoregressive entropy models for
  learned image compression,'' in \emph{2020 IEEE International Conference on
  Image Processing (ICIP)}, 2020, pp. 3339--3343.

\bibitem{He_2022_CVPR}
D.~He, Z.~Yang, W.~Peng, R.~Ma, H.~Qin, and Y.~Wang, ``Elic: Efficient learned
  image compression with unevenly grouped space-channel contextual adaptive
  coding,'' in \emph{Proceedings of the IEEE/CVF Conference on Computer Vision
  and Pattern Recognition (CVPR)}, June 2022, pp. 5718--5727.

\bibitem{He_2021_CVPR}
D.~He, Y.~Zheng, B.~Sun, Y.~Wang, and H.~Qin, ``Checkerboard context model for
  efficient learned image compression,'' in \emph{Proceedings of the IEEE/CVF
  Conference on Computer Vision and Pattern Recognition (CVPR)}, June 2021, pp.
  14\,771--14\,780.

\bibitem{Klopp_2022_Action}
J.~P. Klopp, L.-G. Chen, and S.-Y. Chien, ``Utilising low complexity cnns to
  lift non-local redundancies in video coding,'' \emph{IEEE Transactions on
  Image Processing}, vol.~29, pp. 6372--6385, 2020.

\bibitem{TDAN_CVPR_2020}
Y.~Tian, Y.~Zhang, Y.~Fu, and C.~Xu, ``Tdan: Temporally-deformable alignment
  network for video super-resolution,'' in \emph{2020 IEEE/CVF Conference on
  Computer Vision and Pattern Recognition (CVPR)}, 2020, pp. 3357--3366.

\bibitem{EDVR_CVPR_2019}
X.~Wang, K.~C. Chan, K.~Yu, C.~Dong, and C.~C. Loy, ``Edvr: Video restoration
  with enhanced deformable convolutional networks,'' in \emph{2019 IEEE/CVF
  Conference on Computer Vision and Pattern Recognition Workshops (CVPRW)},
  2019, pp. 1954--1963.

\bibitem{FVC}
Z.~Hu, D.~Xu, G.~Lu, W.~Jiang, W.~Wang, and S.~Liu, ``Fvc: An end-to-end
  framework towards deep video compression in feature space,'' \emph{IEEE
  Transactions on Pattern Analysis and Machine Intelligence}, vol.~45, no.~4,
  pp. 4569--4585, 2023.

\bibitem{Chen_Distillation}
H.~Chen, Y.~Wang, H.~Shu, C.~Wen, C.~Xu, B.~Shi, C.~Xu, and C.~Xu, ``Distilling
  portable generative adversarial networks for image translation,''
  \emph{Proceedings of the AAAI Conference on Artificial Intelligence},
  vol.~34, no.~04, pp. 3585--3592, Apr. 2020.

\bibitem{Yim_2017_CVPR}
J.~Yim, D.~Joo, J.~Bae, and J.~Kim, ``A gift from knowledge distillation: Fast
  optimization, network minimization and transfer learning,'' in
  \emph{Proceedings of the IEEE Conference on Computer Vision and Pattern
  Recognition (CVPR)}, July 2017.

\bibitem{gu2022openvocabulary}
\BIBentryALTinterwordspacing
X.~Gu, T.-Y. Lin, W.~Kuo, and Y.~Cui, ``Open-vocabulary object detection via
  vision and language knowledge distillation,'' in \emph{International
  Conference on Learning Representations}, 2022. [Online]. Available:
  \url{https://openreview.net/forum?id=lL3lnMbR4WU}
\BIBentrySTDinterwordspacing

\bibitem{Li_KD_2022}
S.~Li, M.~Lin, Y.~Wang, Y.~Wu, Y.~Tian, L.~Shao, and R.~Ji, ``Distilling a
  powerful student model via online knowledge distillation,'' \emph{IEEE
  Transactions on Neural Networks and Learning Systems}, pp. 1--10, 2022.

\bibitem{Yu_KD_2023}
Y.~Yu, B.~Li, Z.~Ji, J.~Han, and Z.~Zhang, ``Knowledge distillation classifier
  generation network for zero-shot learning,'' \emph{IEEE Transactions on
  Neural Networks and Learning Systems}, vol.~34, no.~6, pp. 3183--3194, 2023.

\bibitem{Yang_detection_KD}
Z.~Yang, Z.~Li, X.~Jiang, Y.~Gong, Z.~Yuan, D.~Zhao, and C.~Yuan, ``Focal and
  global knowledge distillation for detectors,'' in \emph{2022 IEEE/CVF
  Conference on Computer Vision and Pattern Recognition (CVPR)}, jun 2022, pp.
  4633--4642.

\bibitem{Tung_KD_2019}
F.~Tung and G.~Mori, ``Similarity-preserving knowledge distillation,'' in
  \emph{2019 IEEE/CVF International Conference on Computer Vision (ICCV)},
  2019, pp. 1365--1374.

\bibitem{Heo_KD_2019}
B.~Heo, J.~Kim, S.~Yun, H.~Park, N.~Kwak, and J.~Y. Choi, ``A comprehensive
  overhaul of feature distillation,'' in \emph{2019 IEEE/CVF International
  Conference on Computer Vision (ICCV)}, 2019, pp. 1921--1930.

\bibitem{GAN_distillation_image_compression}
L.~Helminger, R.~Azevedo, A.~Djelouah, M.~Gross, and C.~Schroers,
  ``Microdosing: Knowledge distillation for gan based compression,'' 2022.

\bibitem{Kodak}
\BIBentryALTinterwordspacing
\emph{Kodak PhotoCD dataset, \url{http://r0k.us/graphics/kodak/}}. [Online].
  Available: \url{http://r0k.us/graphics/kodak/}
\BIBentrySTDinterwordspacing

\bibitem{Tecnick}
\BIBentryALTinterwordspacing
\emph{Tecnick dataset, \url{https://bellard.org/bpg/}}. [Online]. Available:
  \url{https://bellard.org/bpg/}
\BIBentrySTDinterwordspacing

\bibitem{CLIC}
\BIBentryALTinterwordspacing
\emph{CLIC dataset, \url{http://www.compression.cc/}}. [Online]. Available:
  \url{http://www.compression.cc/}
\BIBentrySTDinterwordspacing

\bibitem{LIU_dataset}
J.~Liu, D.~Liu, W.~Yang, S.~Xia, X.~Zhang, and Y.~Dai, ``A comprehensive
  benchmark for single image compression artifact reduction,'' \emph{IEEE
  Transactions on Image Processing}, vol.~29, pp. 7845--7860, 2020.

\bibitem{Hu_AAAI}
Y.~Hu, W.~Yang, and J.~Liu, ``Coarse-to-fine hyper-prior modeling for learned
  image compression,'' in \emph{Proceedings of the AAAI Conference on
  Artificial Intelligence}, vol.~34, no.~07, 2020, pp. 11\,013--11\,020.

\bibitem{Cheng2020_paper}
Z.~Cheng, H.~Sun, M.~Takeuchi, and J.~Katto, ``Energy compaction-based image
  compression using convolutional autoencoder,'' \emph{IEEE Transactions on
  Multimedia}, vol.~22, no.~4, pp. 860--873, 2020.

\bibitem{Lee_2020}
J.~Lee, S.~Cho, and S.-K. Beack, ``Context-adaptive entropy model for
  end-to-end optimized image compression,'' in \emph{International Conference
  on Learning Representations}, 2019.

\bibitem{lu2022high}
M.~Lu and Z.~Ma, ``High-efficiency lossy image coding through adaptive
  neighborhood information aggregation,'' \emph{arXiv preprint
  arXiv:2204.11448}, 2022.

\bibitem{Hu_2021}
Y.~Hu, W.~Yang, Z.~Ma, and J.~Liu, ``Learning end-to-end lossy image
  compression: A benchmark,'' \emph{IEEE Transactions on Pattern Analysis and
  Machine Intelligence}, pp. 1--1, 2021.

\bibitem{BDRate}
G.~Bjontegaard, ``Calculation of average {PSNR} differences between {RD}
  curves,'' 2001, {VCEG}-M33.

\bibitem{Sun_2021}
H.~Sun, L.~Yu, and J.~Katto, ``Learned image compression with fixed-point
  arithmetic,'' in \emph{2021 Picture Coding Symposium (PCS)}, 2021, pp. 1--5.

\end{thebibliography}

\end{document}